\documentclass[fleqn,usenatbib]{mnras}

\usepackage[T1]{fontenc}
\usepackage{ae,aecompl}

\usepackage[figurename=Figure.,
                  justification=RaggedRight, 
                  labelfont={bf, footnotesize}, 
                  textfont={footnotesize},position=top]{caption}

\usepackage{graphicx}
\usepackage{amsmath}
\usepackage{amssymb}
\usepackage{float}
\usepackage{dblfloatfix}
\usepackage{blindtext}
\usepackage{booktabs}
\usepackage{rotating}
\usepackage{natbib}
\usepackage{txfonts}
\usepackage{fmtcount}
\usepackage{threeparttablex}
\usepackage{multirow}
\usepackage{morefloats}
\usepackage{dblfloatfix}
\usepackage{dcolumn}
\usepackage{amsmath}
\usepackage{latexsym}
\usepackage{longtable}
\usepackage{lipsum} 
\usepackage{placeins}
\usepackage{textcomp}
\usepackage{gensymb}
\usepackage{mathtools}
\usepackage{afterpage}
\usepackage{caption}
\usepackage{hyperref}

\usepackage{amsmath}
\usepackage{longtable}
\usepackage{rotating}
\usepackage{tikz}
\usepackage{float}
\usepackage{lipsum}

\title[]{SN~2020ank: a bright and fast-evolving H-deficient superluminous supernova}

\author[Kumar et al.]{Amit Kumar$^{1,2}$\thanks{E-mail: amit@aries.res.in},
Brajesh Kumar$^{1,3}$, 
S. B. Pandey$^{1}$\thanks{E-mail: shashi@aries.res.in},
D. K. Sahu$^{3}$,
Avinash Singh$^{1,3}$,
\newauthor G. C. Anupama$^{3}$,
Amar Aryan$^{1,4}$,
Rahul Gupta$^{1,4}$,
Anirban Dutta$^{3}$ and Kuntal Misra$^{1}$\\ \\
$^{1}$Aryabhatta Research Institute of Observational Sciences, Manora Peak, Nainital 263002, Uttarakhand, India\\
$^{2}$School of Studies in Physics and Astrophysics, Pandit Ravishankar Shukla University, Raipur 492010, Chhattisgarh, India\\
$^{3}$Indian Institute of Astrophysics, II Block, Koramangala, Bengaluru 560034, Karnataka, India\\
$^{4}$Department of Physics, Deen Dayal Upadhyaya Gorakhpur University, Gorakhpur 273009, Uttar Pradesh, India
}

\date{Accepted: 2020 December 29; Revised: 2020 December 10; Received: 2020 October 6}

\pubyear{2020}

\begin{document}

\label{firstpage}
\pagerange{\pageref{firstpage}--\pageref{lastpage}}
\maketitle

\begin{abstract}
We investigate the observational properties of a hydrogen-deficient superluminous supernova (SLSN) SN~2020ank (at z = 0.2485), with the help of early phase observations carried out between --21 and +52 d since $g$-band maximum. Photometrically, SN~2020ank is one of the brightest SLSN ($M_{g,peak}$ $\sim$\,--21.84 $\pm$ 0.10 mag), having fast pre-peak rising and post-peak decaying rates. The bolometric light curve of SN~2020ank exhibits a higher peak luminosity ($L_{max}$) of $\sim$\,(3.9 $\pm$ 0.7) $\times$ 10$^{44}$ erg s$^{-1}$ and appears to be symmetric around the peak with $L^{rise}_{max}$/e $\approx$ $L^{fall}_{max}$/e $\approx$ 15 d. The semi-analytical light-curve modelling using the {\tt MINIM} code suggests a spin down millisecond magnetar with $P_i$ $\sim$\,2.2 $\pm$ 0.5 ms and $B$ $\sim$\,(2.9 $\pm$ 0.1) $\times$ $10^{14}$ G as a possible powering source for SN~2020ank. The possible magnetar origin and excess ultraviolet flux at early epochs indicate a central-engine based powering source for SN~2020ank. Near-peak spectra of SN~2020ank are enriched with the W-shaped O\,{\sc ii} features but with the weaker signatures of C\,{\sc ii} and Fe\,{\sc iii}. Using the estimated rise time of $\sim$\,27.9 d and the photospheric velocity of $\sim$\,12050 km s$^{-1}$, we constrain the ejecta mass to $\sim$\,7.2 $M_{\odot}$ and the kinetic energy of $\sim$6.3 $\times$ 10$^{51}$ erg. The near-peak spectrum of SN~2020ank exhibits a close spectral resemblance with that of fast-evolving SN~2010gx. The absorption features of SN~2020ank are blueshifted compared to Gaia16apd, suggesting a higher expansion velocity. The spectral similarity with SN~2010gx and comparatively faster spectral evolution than PTF12dam (a slow-evolving SLSN) indicate the fast-evolving behavior of SN~2020ank.

\end{abstract}


\section{Introduction} \label{sec:Intro}

Superluminous supernovae (SLSNe) are nearly 2--3 magnitudes brighter than classical SNe \citep{Angus2019, Gal-Yam2019a, Inserra2019} radiating total energy of the order of $\sim$10$^{51}$ erg and exhibit characteristic W-shaped O\,{\sc ii} features towards blue in the near-peak spectra \citep{Quimby2011, Quimby2018, Gal-Yam2019}. SLSNe are rare class of events with high-peak luminosity and were unknown before SN~2005ap \citep{Quimby2007}. They comprise $\sim$0.01\% of normal core-collapse SNe (CCSNe), and nearly 150 objects have been spectroscopically confirmed so far \citep{Quimby2013, McCrum2015, Liu2017a, Prajs2017, Gomez2020}. Based on the Hydrogen abundance, these events are broadly classified into H-poor SLSNe (SLSNe~I) and H-rich SLSNe (SLSNe~II;  \citealt{Gal-Yam2012, Branch2017}). Most of the SLSNe~I generally occur in metal-poor faint dwarf galaxies \citep{Lunnan2014,Chen2017b} with complex light curves having pre-peak bumps (e.g., LSQ14bdq; \citealt{Nicholl2015b}; see also \citealt{Angus2019}) and post-peak undulations (e.g., SN~2015bn; \citealt{Nicholl2016a}). On the other hand, SLSNe~II mostly present prominent and narrow hydrogen Balmer lines, also characterized as SLSNe~IIn due to their spectral similarity with lower luminosity SNe~IIn (e.g., SN~2008am; \citealt{Chatzopoulos2011}). However, some of these events lack the typical narrow hydrogen features, e.g., SN~2008es \citep{Gezari2009, Miller2009}, SN~2013hx and PS15br \citep{Inserra2018b}. SLSNe~IIn have primarily been found in heterogeneous host environments and remain less-studied \citep{Leloudas2015, Perley2016, Schulze2018}.

SLSNe~I appear to have slow- and fast-evolving behaviour based on their different photometric and spectroscopic properties \citep{Inserra2017, Vreeswijk2017, Pursiainen2018, Quimby2018, Inserra2018a, Inserra2019, Reka2020b}. Photometrically slow-evolving (``PTF12dam-like''; rise time $\sim$33-100 d) SLSNe~I also exhibit slower spectroscopic evolution in comparison to the fast-evolving (``SN~2011ke-like''; rise time $\sim$13-35 d) SLSNe~I \citep{Quimby2018}. In addition, the slow-evolving SLSNe~I have lower SN~expansion velocity ($v_{\rm exp}$) of $\lesssim$12000 km s$^{-1}$ and shallower velocity gradient (between 10 and 30 d, post-peak), whereas fast-evolving ones show comparatively higher $v_{\rm exp}$ of $\gtrsim$12000 km s$^{-1}$ and steeper velocity gradient in the same time regime \citep{Inserra2018b}.

The physical mechanism giving rise to the high peak-luminosity feature in most of the SLSNe~I remains debatable. The widely accepted physical mechanism of radioactive decay (RD) of $^{56}$Ni for normal class of H-deficient CCSNe has been found to be inefficient in explaining the observed high peak-luminosity in most of the SLSNe~I. Theoretically, this would require a higher nickel mass ($M_{Ni}$) synthesis ($\gtrsim 5\,M_\odot$; \citealt{Gal-Yam2012}), generally not possible in the core-collapse system \citep{Umeda2002, Umeda2008}. An alternate theory based on pair-instability SNe \citep[PISNe;][]{Kozyreva2015} has been considered to explain features in some of the slow-evolving SLSNe~I (e.g., SN~2007bi; \citealt{Gal-Yam2009}). However, the sharper pre-peak rising rates and bluer colours of these objects are not favorable with this scenario \citep{Kasen2011, Dessart2012, Jerkstrand2017}. Various other plausible models are also proposed to explain the relatively wider and luminous bolometric light curves of these ultraviolet (UV) bright cosmic events, including Circumstellar Matter Interaction \citep[CSMI;][]{Ginzburg2012, Wheeler2017}, spin-down Millisecond Magnetar \citep[MAG;][]{Kasen2010,Woosley2010,Metzger2015,Chen2017a,Dessart2019, Lin2020ApJ}, and their possible combinations, termed as ``HYBRID'' \citep{Chatzopoulos2012} models, e.g., CSMI + RD, CSMI + MAG, CSMI + RD + MAG \citep{Chatzopoulos2013, Moriya2018, Chatzopoulos2019, Wang2019}.

Detailed studies for individual SLSNe~I reveal that no single model is sufficient to explain all observed phenomena in these systems. For example, a handful of SLSNe~I (iPTF13ehe; \citealt{Yan2015}, iPTF15esb and iPTF16bad; \citealt{Yan2017b}) manifest clear spectral signatures supporting the CSMI. The shock-cooling of the extended CSM usually explains the observed pre-peak bumps in the light-curves of SLSNe~I (e.g., SN~2006oz; \citealt{Leloudas2012}, DES14X3taz; \citealt{Smith2016}; see also \citealt{Piro2015}). Whereas ejecta interaction with the pre-expelled CSM shells is considered as the potential reason for the post-peak undulations (e.g., SN~2007bi; \citealt{Gal-Yam2009}, SN~2015bn; \citealt{Nicholl2016a}), favouring the CSMI. On the other hand, there are a few observational features that favor the MAG model. For instance, the near-peak excess $UV$ flux in the case of Gaia16apd \citep{Nicholl2017} is explained in terms of a central engine based power source: it may be a spin-down millisecond magnetar or a mass accreting black-hole \citep[][]{MacFadyen1999}. Similarly, SLSN~2011kl, the only known case so far associated with the ultra-long Gamma-Ray Burst (Ul-GRB), e.g., GRB 111209A \citep{Greiner2015}, also supports the central engine based powering mechanism \citep{Bersten2016ApJ, Lin2020} and hints that some of these SLSNe~I may also be connected with long GRBs \citep{Kann2019}. We note that both the models (CSMI and MAG) can explain various observational aspects in the SLSNe~I light curves \citep{Inserra2013, Nicholl2014}, though a few specific features favour the MAG model, including the near-peak high $UV$ flux \citep{Mazzali2016, Nicholl2017}, late-time flattening \citep{Inserra2013, Liu2017b, Nicholl2017a, Blanchard2018}, and spectral properties \citep{Dessart2012, Nicholl2019}. Though deeper investigations are required to explore the underlying physical mechanisms, possible progenitors and environments hosting such rare and energetic explosions.

SN~2020ank (ZTF20aahbfmf) was discovered by the Zwicky Transient Facility \citep[ZTF; ][]{Bellm2019} on 2020 January 19 at J2000 coordinates: RA = $08^{\rm h}16^{\rm m}14\fs65$ and Dec = $+04\degr 19\arcmin 26\farcs87$ \citep{Poidevin2020a}. SN~2020ank was also detected by the Asteroid Terrestrial-impact Last Alert System \citep[ATLAS;][]{Tonry2018} with internal name ATLAS20dzr on 2020 January 24 \citep{Tonry2020} and by the Pan-STARRS1 \citep[PS1;][]{Chornock2013} on 2020 March 18 as PS20eyd. SN~2020ank was classified as an SLSN~I based on the spectroscopic observations from the Liverpool Telescope (LT-2.0m) and the Gran Telescopio Canarias \citep[GTC-10.4m;][]{Poidevin2020b}. Later on, spectroscopic observations were also carried out by \citet{Dahiwale2020}, discussing the spectrum obtained from the Palomar Observatory Hale Telescope (P200-5.1m) reporting redshift z = 0.2485. The near-peak polarimetric observations showing negligible polarization, as investigated by \citet{Lee2020}, suggests a nearly spherical explosion for SN~2020ank.

In this paper, early time photometric and spectroscopic observations of SN~2020ank have been discussed. The paper is structured as follows. The procedures describing observations, data reductions, and analysis are explained in Section~\ref{sec:reduction}. In Section~\ref{sec:light_curves}, the photometric properties of SN~2020ank and its comparison with other well-studied SLSNe~I are presented. The bolometric light-curve modelling using the {\tt MINIM} code is presented in Section~\ref{sec:bolo}. Section~\ref{sec:spectra} describes the spectroscopic properties of SN~2020ank, the {\tt SYNAPPS} spectral modelling, and the spectral comparison with other well-studied SLSNe~I. We conclude our results in Section~\ref{sec:results}. Throughout this work, H$_0$ = 70 km s$^{-1}$ Mpc$^{-1}$ and $\Omega_m$ = 0.27 have been adopted to estimate the distances, dates are presented in UT, and phase is given since $g$-band maximum.

\begin{figure*}
\includegraphics[angle=0,scale=0.275]{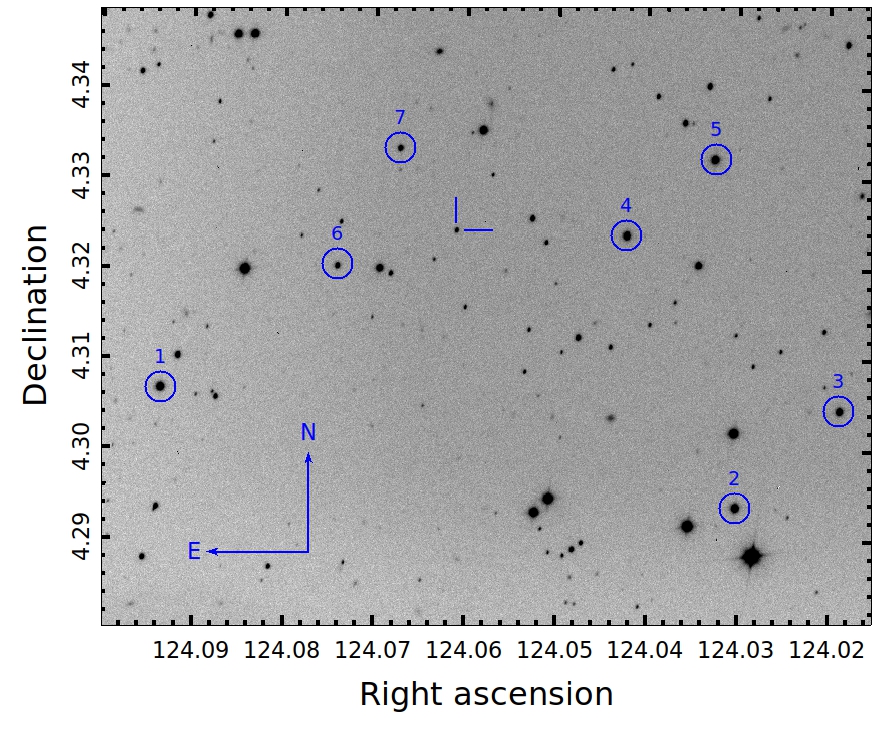}
\includegraphics[angle=0,scale=0.70]{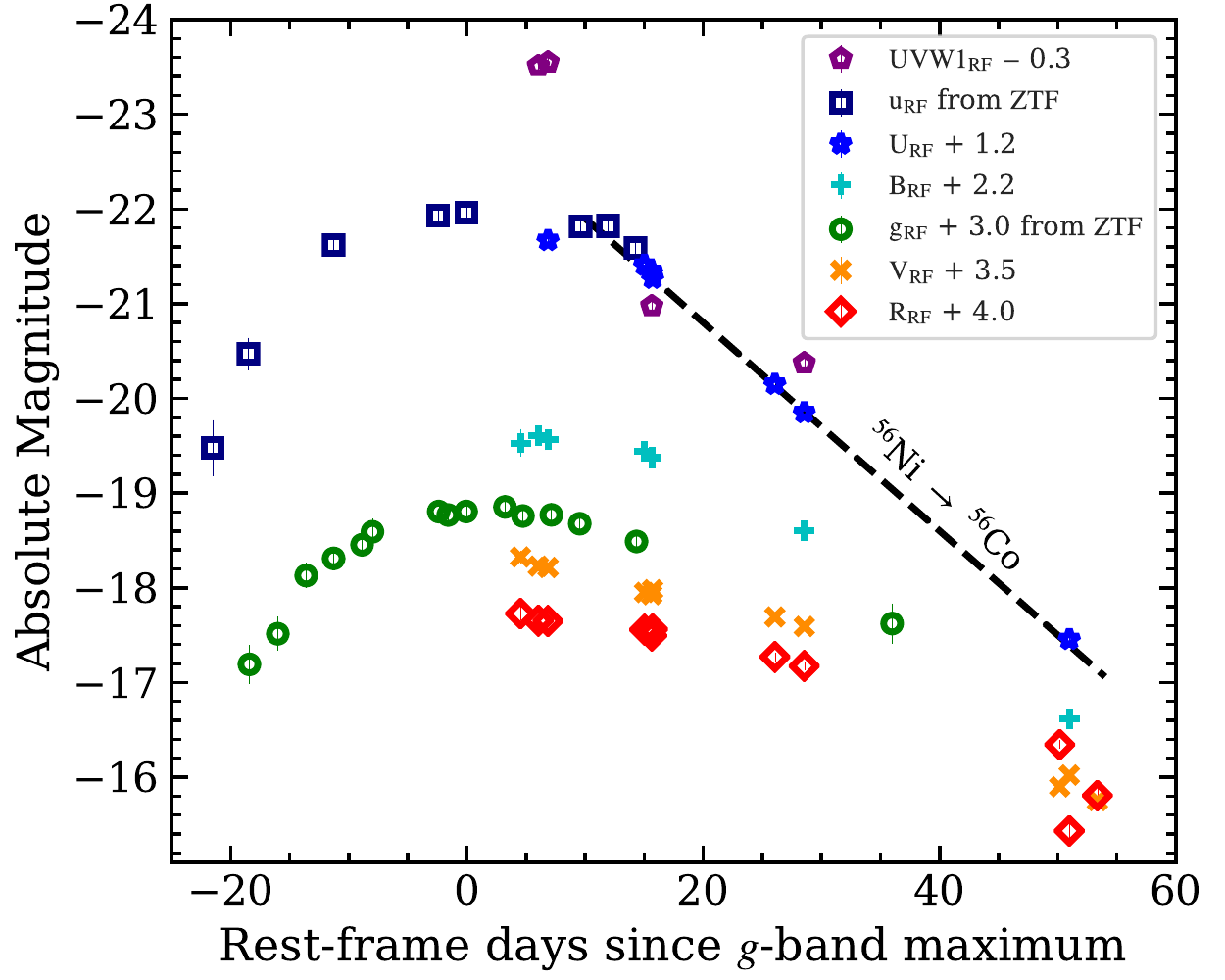}
\caption{$Left$: Finding chart of SN~2020ank ($\sim$\,5 arcmin $\times$ 4 arcmin~field) along with other local standard stars (marked in circles, IDs 1--7). The $R$-band image (exposure time = 5 min) observed on 2020 March 19 using the 4K $\times$ 4K CCD Imager mounted at the axial port of the DOT-3.6m. The location of SN is marked with a blue crosshair. North and East directions are also indicated in the image. $Right$: The temporal evolution of SN~2020ank in $UVW1_{RF}$, $u_{RF}$, $U_{RF}$, $B_{RF}$, $g_{RF}$, $V_{RF}$, and $R_{RF}$ bands. The light curves are corrected for the Galactic extinction and the $K$-corrections are applied using our spectral analysis. The post-peak decay rate of $U_{RF}$ light curve is well in agreement with the decay rate of $^{56}Ni$ $\rightarrow$ $^{56}Co$ theoretical curve (0.11 mag d$^{-1}$).}
\label{fig:phot}
\end{figure*}

\section{Observations and data analysis}\label{sec:reduction}
Photometric observations in Bessell $U$, $B$, $V$, $R$, and $I$ bands of SN~2020ank field were carried out with three ground-based observing facilities in India: Sampurnanand Telescope (ST-1.04m), Himalayan Chandra Telescope (HCT-2.0m), and recently commissioned Devasthal Optical Telescope (DOT-3.6m) having longitudinal advantage for time critical observations \citep{Pandey2016, Pandeyshashi2018}. These three telescopes are equipped with liquid nitrogen cooled CCD cameras at their Cassegrain focus. The observations were initiated using the ST-1.04m and continued with the HCT-2.0m (4 epochs) and DOT-3.6m (6 epochs). Photometric images were acquired with the 4K $\times$ 4K CCD Imagers mounted at the axial ports of both the ST-1.04m and the DOT-3.6m \citep{Pandey2018}. In addition, the Himalayan Faint Object Spectrograph and Camera (HFOSC\footnote{\url{https://www.iiap.res.in/?q=iao_about}}) instrument at the HCT-2.0m has also been used to perform the multiband photometric data of SN~2020ank. Table~\ref{inst} lists various parameters of the facilities and their back-end instruments. Standard $IRAF$\footnote{\url{http://iraf.noao.edu/}} tasks were executed to pre-process (e.g., bias-subtraction, flat-fielding, and cosmic ray removal) the raw data. On several nights, multiple science frames in each band were stacked after the alignment of the individual images, and consequently, a better signal-to-noise ratio was obtained. To calibrate a sequence of secondary standards in the SN field, we observed Landolt photometric standard fields PG~1657 and PG~0231 \citep{Landolt1992}. The standard and SN fields were observed on 2020 March 19 (PG~1657) and 2020 October 14 (PG~0231) with the DOT-3.6m under good photometric conditions. The Landolt field stars have a brightness range of 12.77\,$\le$\,$V$\,$\le$16.11 mag and colour range of --\,0.15\,$\le$\,$B-V$\,$\le$ 1.45 mag. Using the stand-alone version of DAOPHOT\footnote{{\sc DAOPHOT} stands for Dominion Astrophysical Observatory Photometry.} \citep{Stetson1987, Stetson1992}, the point spread function photometry on all the frames was performed. The average atmospheric extinction values in $U$, $B$, $V$, $R$, and $I$ bands for the Devasthal site were adopted from \citet{Mohan1999}. Using the Landolt standards, transformation to the standard system was derived by applying average colour terms, and photometric zero-points. In the left-hand panel of Fig.~\ref{fig:phot}, the seven secondary calibrated standard stars (used to calibrate the SN magnitudes) are marked, and the respective $U$, $B$, $V$, $R$, and $I$ magnitudes are listed in Table~\ref{tab_id}. The final SN photometry in $U$, $B$, $V$, $R$, and $I$ bands are listed in Table~\ref{tab:photodata}. Here, we note that for completeness, in addition to our observations, the publicly available $ZTF$ $g$ and $r$ bands data are also used \citep{Poidevin2020a} and downloaded from the Lasair\footnote{\url{https://lasair.roe.ac.uk/object/ZTF20aahbfmf/}} website \citep{Smith2019}.

The optical spectroscopic observations of SN~2020ank in low-resolution mode were performed at four epochs from HCT-2.0m using the HFOSC instrument. The spectra were obtained using grism-Gr7 (3500--7800 \AA), having a resolution of $\sim$8 \AA. The journal of these observations is provided in Table~\ref{tab:tablespec}. The arc lamp and spectrophotometric standards were also obtained during the observations, and the spectroscopic data reduction was performed using the \textit{IRAF} software. The pre-processing of raw spectra, extraction of 1D spectra, and the wavelength calibration were done in a standard manner as described in \cite{Kumar2018}. For flux calibration, spectrophotometric standard observations were used. The flux calibrated spectra were then scaled with respect to the calibrated photometric $U$, $B$, $V$, $R$, and $I$ fluxes to bring them to an absolute flux scale and, finally, corrected for the host galaxy redshift. In this study, we have also used two early epoch archival spectra obtained using the OSIRIS\footnote{Optical System for Imaging and low Resolution Integrated Spectroscopy.} at GTC-10.4m \citep{Poidevin2020b} and DBSP\footnote{Double Spectrograph.} at P200-5.1m \citep{Dahiwale2020} and downloaded from the Transient Name Server (TNS)\footnote{\url{https://wis-tns.weizmann.ac.il/}}.

\begin{table*}
\small{
 \begin{center}
  \begin{threeparttable}
    \caption{Observing facilities and instruments detail used for the follow-up observations of SN~2020ank.}
    \label{inst}
    \addtolength{\tabcolsep}{-2pt}
    \begin{tabular}{llcccccc} 
\hline
Facility                          &  Location             &   Instrument        & Gain          & Readout Noise &  Binning & Plate scale$^a$ & Field of view\\ 
$ $                               &  $ $                  &   $ $               & (e$^{-}/ADU$) &  (e$^{-}$) &    & (arcsec pixel$^{-1}$) & (arcmin$^2$) \\ \hline  \hline
1.04-m Sampurnanand Telescope     & Manora Peak, Nainital & 4K$\times$4K Imager & 3.0           &      10.0  &  4 $\times$ 4   & 0.230  & 15 $\times$ 15 \\
2.0-m Himalayan Chandra Telescope & Hanle, Leh            & HFOSC               & 0.28          &      5.75  &  1 $\times$ 1   & 0.296  & 10 $\times$ 10 \\
3.6-m Devasthal Optical Telescope & Devasthal, Nainital   & 4K$\times$4K Imager & 5.0           &      10.0  &  2 $\times$ 2   & 0.095  & 6.5 $\times$ 6.5 \\
\hline
       \hline 
    \end{tabular}
 \begin{tablenotes}[para,flushleft]
        $^a$ Plate scales are given for un-binned mode.
    \end{tablenotes}

  \end{threeparttable}
 \end{center}}
\end{table*}

\begin{table*}
\centering
\small{
\caption{Identification number (ID) and calibrated magnitudes of the secondary standard stars in the field of SN~2020ank as displayed in the left-hand panel of Fig.~\ref{fig:phot}.}
% \scriptsize
\addtolength{\tabcolsep}{14pt}
\label{tab_id}
\begin{tabular}{cccccc}
\hline
Star &    $U$             &   $B$              &   $V$               &   $R$              &   $I$              \\
ID   &   (mag)            &  (mag)             &  (mag)              &  (mag)             &  (mag)             \\ \hline \hline
1    & 17.749 $\pm$ 0.010 & 17.353 $\pm$ 0.005 & 16.615 $\pm$ 0.003  & 16.152 $\pm$ 0.005 & 15.770 $\pm$ 0.008 \\
2    & 17.853 $\pm$ 0.017 & 17.450 $\pm$ 0.005 & 16.662 $\pm$ 0.006  & 16.229 $\pm$ 0.007 & 15.854 $\pm$ 0.008 \\
3    & 17.747 $\pm$ 0.010 & 17.765 $\pm$ 0.007 & 17.208 $\pm$ 0.009  & 16.879 $\pm$ 0.008 & 16.579 $\pm$ 0.009 \\
4    & 18.226 $\pm$ 0.010 & 17.798 $\pm$ 0.004 & 16.953 $\pm$ 0.007  & 16.471 $\pm$ 0.006 & 16.008 $\pm$ 0.007 \\
5    & 17.123 $\pm$ 0.009 & 17.069 $\pm$ 0.005 & 16.512 $\pm$ 0.006  & 16.205 $\pm$ 0.004 & 15.909 $\pm$ 0.006 \\
6    & --                 & 18.918 $\pm$ 0.006 & 18.405 $\pm$ 0.005  & 18.095 $\pm$ 0.010 & 17.796 $\pm$ 0.009 \\
7    & --                 & 18.985 $\pm$ 0.011 & 18.302 $\pm$ 0.004  & 17.922 $\pm$ 0.007 & 17.569 $\pm$ 0.011 \\
\hline
\end{tabular}}
\end{table*}

\begin{figure*}
\includegraphics[angle=0,scale=1.0]{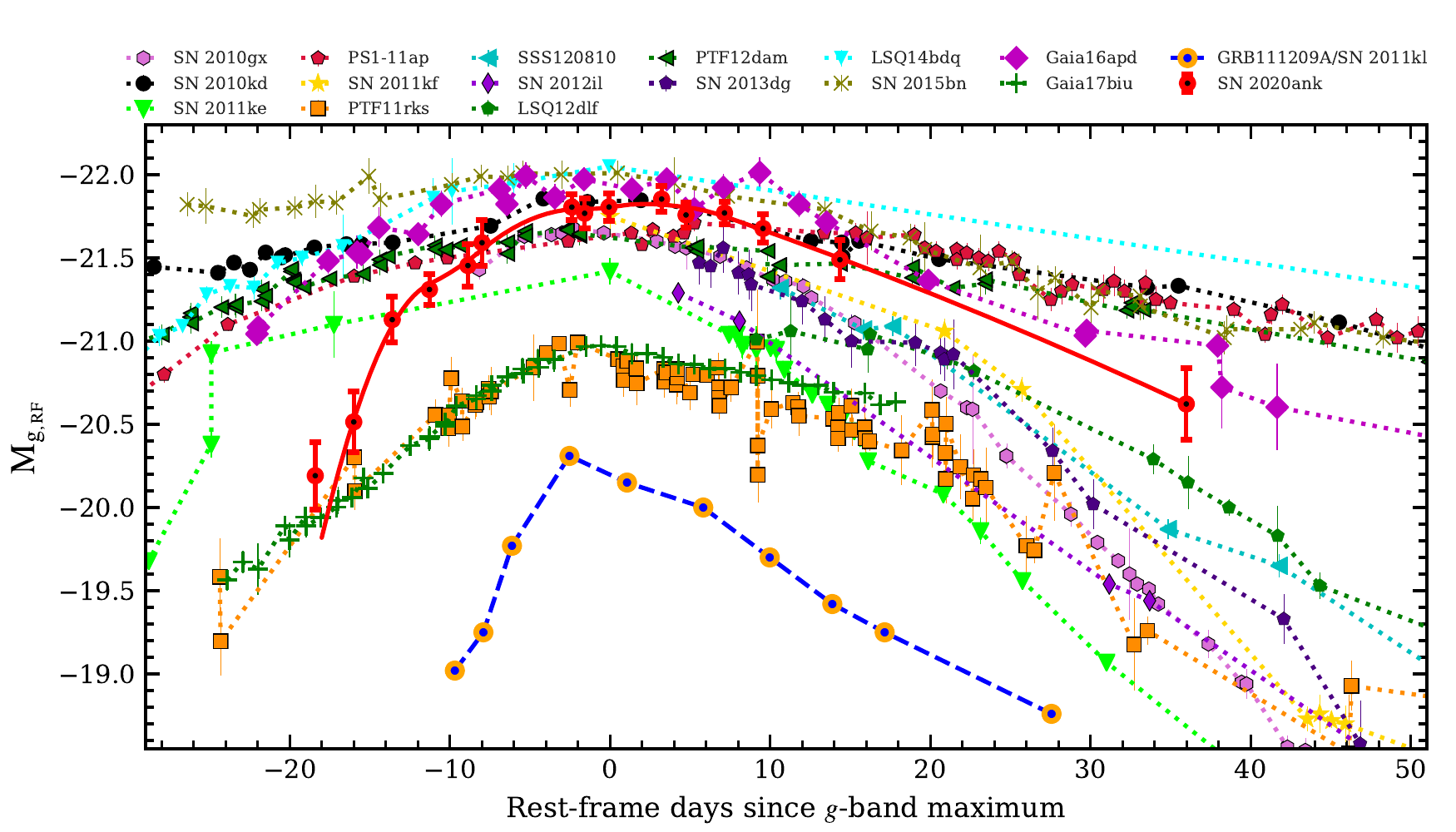}
\caption{Comparison of the rest-frame $M_g$ light curve of SN~2020ank with other well-studied SLSNe~I (taken from \citealt{Nicholl2015a,Nicholl2016a,DeCia2018,Kangas2017,Bose2018,Kumar2020}, and references therein) and the long GRB connected SLSN~2011kl \citep{Greiner2015,Kann2019}. All the presented light curves are corrected for the Galactic extinction, and the $K$-corrections also have been applied. SN~2020ank appears to have comparatively brighter $M_{g,peak}$ and faster pre-peak rising and post-peak decaying rates, apparently similar to SN~2010gx and other fast-evolving SLSNe~I.}
\label{fig:Mg}
\end{figure*}

\begin{figure}
\includegraphics[angle=0,scale=0.55]{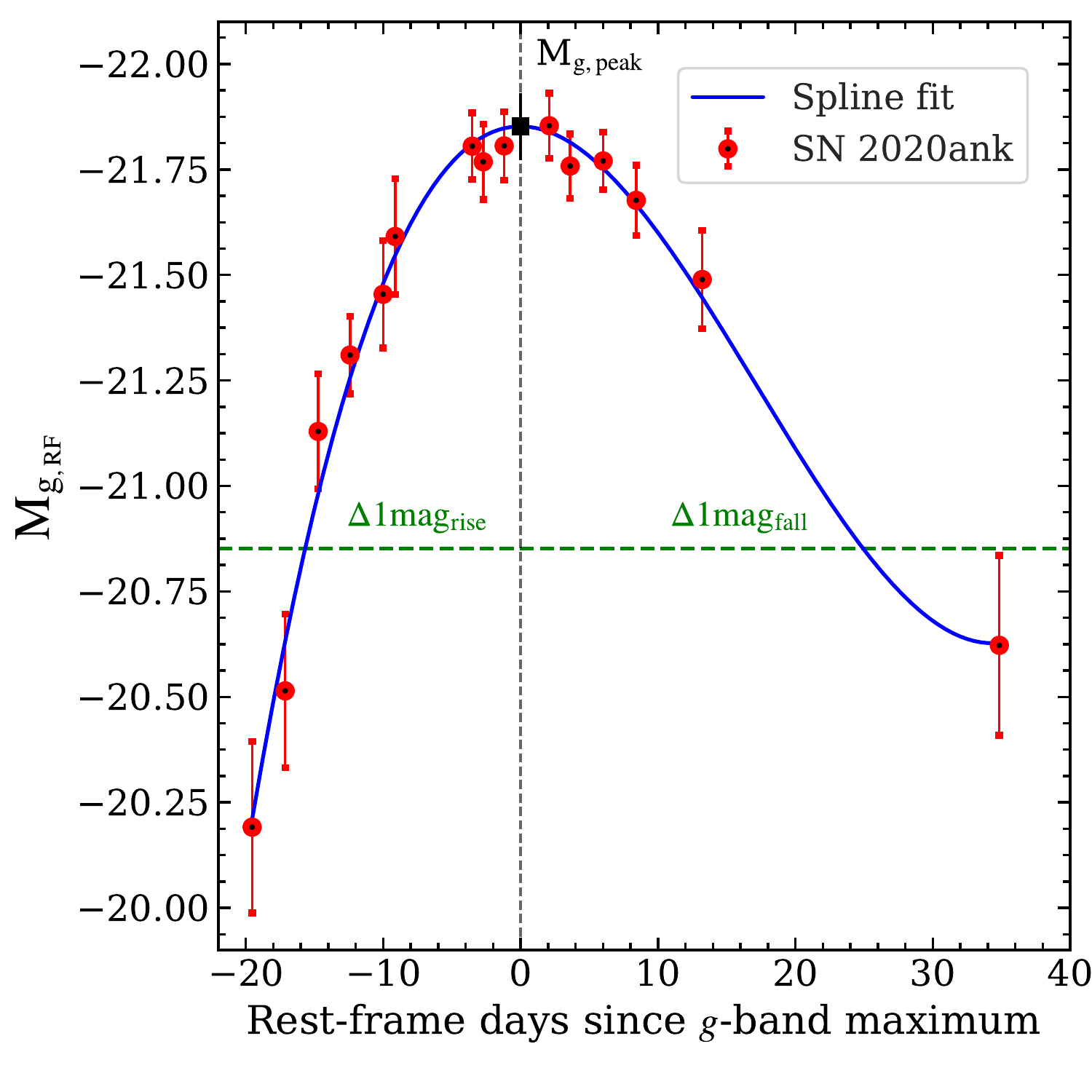}
\caption{The rest-frame $M_g$ light curve of SN~2020ank fitted with the third order spline function to calculate the $M_{g,peak}$, $t^{\Delta 1mag}_{rise}$ and $t^{\Delta 1mag}_{fall}$ is shown. The black square shows the estimated $M_{g,peak}$, whereas the intersect points of green dotted line and spline fit (in blue) present magnitude values constraining $t^{\Delta 1mag}_{rise}$ and $t^{\Delta 1mag}_{fall}$ (in rest-frame d).}
\label{fig:Mg_spline}
\end{figure}

\begin{figure*}
\includegraphics[angle=0,scale=0.90]{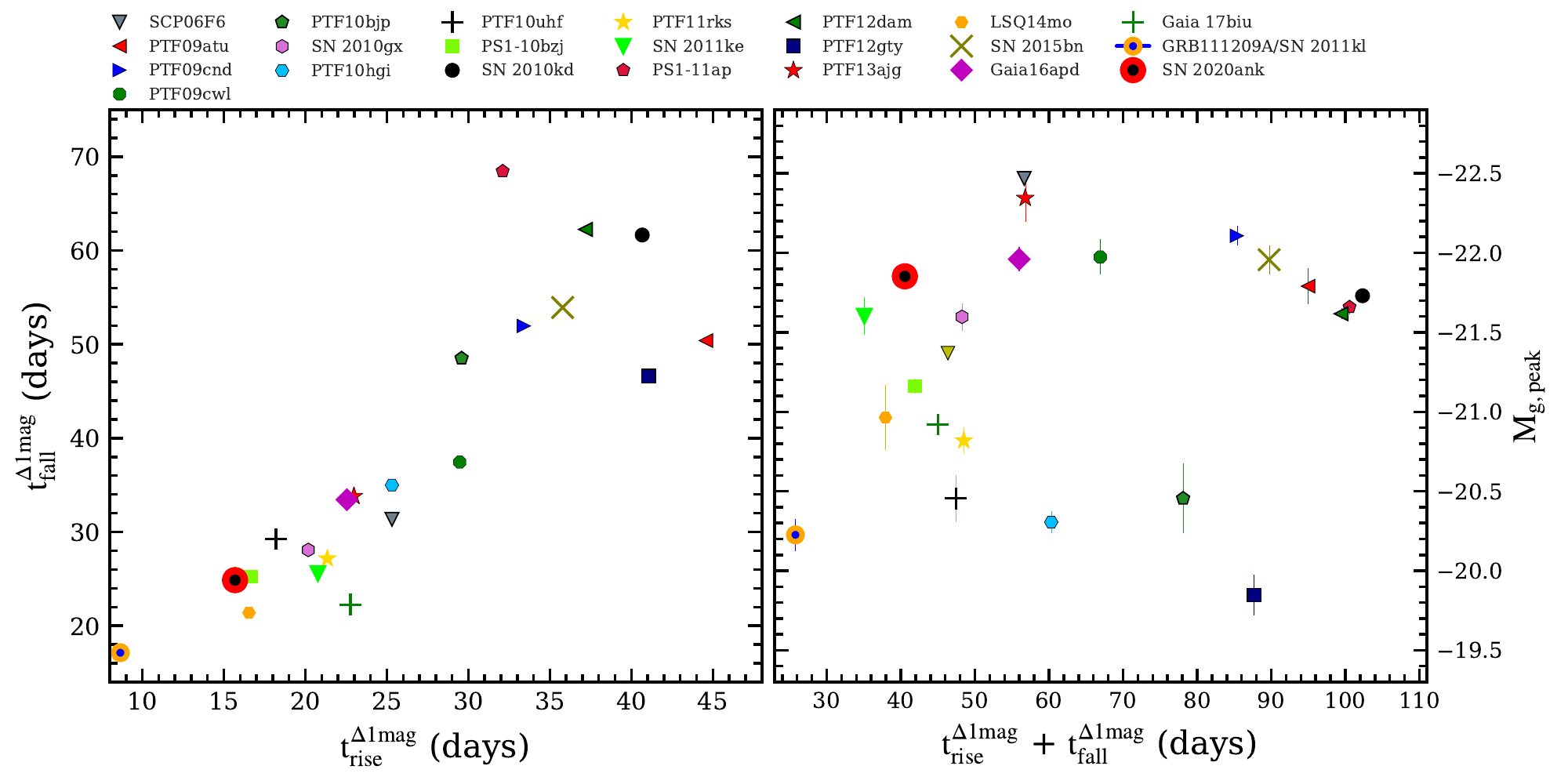}
\caption{$Left$: Estimated values of $t^{\Delta 1mag}_{fall}$ compared with $t^{\Delta 1mag}_{rise}$ indicating nearly a linear relation between the two observed quantities for the set of SLSNe~I. $Right$: The sum of $t^{\Delta 1mag}_{fall}$ and $t^{\Delta 1mag}_{rise}$ compared with $M_{g,peak}$ does not show any such correlation. However, from these plots, it is clear that SN~2020ank is a fast-evolving SLSN with a comparatively brighter peak. On the other hand, the GRB connected SLSN~2011kl \citep{Greiner2015,Kann2019} seems to have the fastest pre-peak rising and post-peak decaying rates in comparison to all other presented SLSNe~I.}
\label{fig:trise_fall_Mg}
\end{figure*}

\section{Light-curve evolution and comparison with other events}
\label{sec:light_curves}

SN~2020ank was discovered on 2020 January 19 UT 09:15:13 (MJD = 58,867.386) in the observed-frame $g$-band at $\sim$20.91 $\pm$ 0.30 mag by the $ZTF$ \citep{Poidevin2020a}. Whereas the last non-detection was on 2020 January 19 UT 08:09:24 (MJD = 58,867.340) also reported by the $ZTF$ with a $r$-band upper limit of $\sim$20.40 mag. The host galaxy of SN~2020ank is faint ($g$-band upper limit $\sim$24 mag) as it was not observed up to the detection limits of Sloan Digital Sky Survey (SDSS), PS1, and Dark Energy Spectroscopic
Instrument Legacy Survey \citep{Poidevin2020b}. Using this brightness limit, we constrain the explosion date (MJD$_{\rm expl}) \approx 58857.9$ by extrapolating the pre-maximum rest frame $g$-band light curve of SN~2020ank down to the limiting magnitude of the host galaxy by fitting a high-order spline function, as also described in \cite{Gezari2009, Kumar2020}. Our observations of SN~2020ank in the Bessell $U$, $B$, $V$, $R$, and $I$ bands span +2.8 to +51.6 d (in the rest-frame). Log of the photometric observations is tabulated in Table~\ref{tab:photodata}. For a better sampling, we also used the publicly available photometric SDSS $g$ and $r$ bands data (between $\sim -$21 and +36 d, from rest-frame) observed by the $ZTF$ \citep{Poidevin2020a}. 
 
The occurrence of SN~2020ank at z = 0.2485 \citep{Dahiwale2020} necessitated the use of applying $K$-corrections to obtain the rest-frame magnitudes. We estimated $K$-corrections using the optical spectra of SN~2020ank with help of the light version of SuperNova Algorithm for $K$-correction Evaluation code \citep[{\tt SNAKELOOP};][]{Inserra2018b} based on the equation given by \citet{Hogg2002}. The spectra had been flux calibrated by scaling them to the photometric data before being used as an input to the {\tt SNAKELOOP} code. The estimated $K$-correction terms using the near peak-spectrum (at $-$0.74 d, in the rest-frame) of SN~2020ank for $U$, $B$, $g$, $V$, $r$, $R$, and $I$ bands are tabulated in Table~\ref{tab:k_corr}. The $K$-correction terms for intervening photometric epochs are obtained using the interpolation technique. However, for the photometric data points outside the spectral range, we used the same $K$-correction term obtained for the last available spectral epoch. Using the $K$-correction magnitude values, we converted $U$ $\rightarrow$ $UVW1_{RF}$, $B$ $\rightarrow$ $U_{RF}$, $g$ $\rightarrow$ $u_{RF}$, $V$ $\rightarrow$ $B_{RF}$, $r$ $\rightarrow$ $g_{RF}$, $R$ $\rightarrow$ $V_{RF}$ and $I$ $\rightarrow$ $R_{RF}$ bands by applying the formula,

\begin{equation}\label{eq:k_corr}
M_{RF} = m_O - (5\, \times \log d_L + 25) - A_O - K_{OR},
\end{equation}

where $M_{RF}$ represents the absolute magnitude in the rest-frame band, $m_O$ is the apparent magnitude in the observed-frame band, $d_L$ is the luminosity distance, $A_O$ is the extinction for the observed band, and $K_{OR}$ is the $K$-correction term estimated from the spectra of SN~2020ank using the {\tt SNAKELOOP} code. The data have been corrected for the Galactic extinction using $E(B-V)$ = 0.019 mag \citep{Schlafly2011}. The weak host emission lines \citep{Osterbrock1989} and insignificant Na~I absorption \citep{Poznanski2012} in near-peak spectra of SN~2020ank implied negligible host extinction.

For SN~2020ank, the multiband ($UVW1_{RF}$, $u_{RF}$, $U_{RF}$, $B_{RF}$, $g_{RF}$, $V$, and $R_{RF}$) rest-frame absolute magnitudes (after extinction and $K$-correction) estimated using equation~\ref{eq:k_corr} are plotted in the right-hand panel of Fig.~\ref{fig:phot}. The phases after correcting for time dilation are plotted with respect to the $g$-band maximum. From here onward, the discussed bands and the phase will be referred to the rest frame. A low-order polynomial was fitted around the approximate rest-frame $g$-band peak brightness to estimate the date of $g$-band maximum. The date of maximum and the peak magnitude are estimated as MJD$_{g,peak}$ $\approx$ 58,894.28 $\pm$ 0.15 and M$_{g,peak}$ $\approx -21.84 \pm 0.10$ mag, respectively. In the $u$-band, SN~2020ank reached a peak brightness of M$_{u,peak}$ $\approx$ $-$21.95 $\pm$ 0.07 mag on MJD$_{u,peak}$ $\approx$ 58,893.36 $\pm$ 0.20. The $UV$ bands appear to peak earlier compared to the redder bands, as found in other CCSNe \citep{Taddia2018}. For SN~2020ank, the M$_{g,peak}$ is consistent with the range of peak absolute magnitudes typically found in the case of other well-studied SLSNe~I \citep{Quimby2013, Nicholl2016a, Inserra2018c}. We observed the post-peak light-curve evolution of SN~2020ank in $UVW1$, $U$, $B$, $V$, and $R$ bands and calculated the decay rates. The redder bands seem to have shallower post-peak decay rates in comparison to the $UV$ bands, which is $\sim$0.05 and $\sim$0.15  mag d$^{-1}$ for the $R$ and $UVW1$ bands, respectively. In the case of SN~2020ank, the redder bands ($V$ and $R$; 0.06 and 0.05 mag d$^{-1}$, respectively) exhibit a steeper post-peak decay rate in comparison to the average $g$-band post-peak decay rate (for $<$ 60 d, 0.04 mag d$^{-1}$) for the SLSNe~I discussed by \citet{DeCia2018}, indicating the comparatively fast-decaying behavior of SN~2020ank. Additionally, the $U$-band post-peak decay rate of SN~2020ank is well in agreement with the $^{56}Ni$ $\rightarrow$ $^{56}Co$ theoretical decay curve (0.11 mag d$^{-1}$), see right-hand panel of Fig.~\ref{fig:phot}.

\begin{table}
\small{
 \begin{center}
    \caption{Optical-photometric data of SN~2020ank in Bessell \textit{U}, \textit{B}, \textit{V}, \textit{R} and \textit{I} bands observed using the ST-1.04, HCT-2.0, and DOT-3.6m. The values presented here are not corrected for Galactic extinction and $K$-corrections. It is also to be noted that the source could not be observed beyond 2020 April 19 due to the unforeseen situation of COVID-19.}
    \label{tab:photodata}
    \addtolength{\tabcolsep}{-2pt}
    \begin{tabular}{cccccc}
\hline
JD & Date & mag & error & Filter & Telescope \\
     $ $  &  &   &  &  \\         
    \hline
   \hline
2458900.117   & 2020  February  20  &    17.879 &  0.041           &   $U$  & HCT-2.0m \\
2458901.133   & 2020  February  21  &    17.886 &  0.072           &   $U$  & HCT-2.0m \\
2458912.094   & 2020  March     16  &    20.501 &  0.062           &   $U$  & DOT-3.6m \\
2458928.209   & 2020  March     19  &    20.856 &  0.071           &   $U$  & DOT-3.6m \\
\hline			  
2458901.133   & 2020  February  21  &    18.455 &  0.035           &   $B$  & HCT-2.0m \\
2458911.343   & 2020  March     02  &    18.930 &  0.028           &   $B$  & HCT-2.0m \\
2458912.094   & 2020  March     03  &    18.998 &  0.021           &   $B$  & HCT-2.0m \\
2458912.193   & 2020  March     03  &    19.069 &  0.030           &   $B$  & DOT-3.6m \\
2458925.109   & 2020  March     16  &    20.134 &  0.021           &   $B$  & DOT-3.6m \\
2458928.209   & 2020  March     19  &    20.402 &  0.023           &   $B$  & DOT-3.6m \\
2458956.212   & 2020  April     16  &    22.797 &  0.093           &   $B$  & DOT-3.6m \\
\hline			  
2458898.229   & 2020  February  18  &    18.542 &  0.149           &   $V$  & ST-1.04m  \\
2458900.117   & 2020  February  20  &    18.466 &  0.021           &   $V$  & HCT-2.0m \\
2458901.133   & 2020  February  21  &    18.526 &  0.020           &   $V$  & HCT-2.0m \\
2458911.343   & 2020  March     02  &    18.690 &  0.023           &   $V$  & HCT-2.0m \\
2458912.094   & 2020  March     03  &    18.753 &  0.021           &   $V$  & HCT-2.0m \\
2458912.094   & 2020  March     16  &    18.758 &  0.019           &   $V$  & DOT-3.6m \\
2458928.209   & 2020  March     19  &    19.502 &  0.020           &   $V$  & DOT-3.6m \\
2458956.212   & 2020  April     16  &    21.489 &  0.035           &   $V$  & DOT-3.6m \\
\hline			  
2458898.229   & 2020  February  18  &    18.307 &  0.070           &   $R$  & ST-1.04m  \\
2458900.117   & 2020  February  20  &    18.389 &  0.019           &   $R$  & HCT-2.0m \\
2458901.133   & 2020  February  21  &    18.373 &  0.029           &   $R$  & HCT-2.0m \\
2458911.343   & 2020  March     02  &    18.611 &  0.026           &   $R$  & HCT-2.0m \\
2458912.094   & 2020  March     03  &    18.642 &  0.029           &   $R$  & HCT-2.0m \\
2458912.193   & 2020  March     03  &    18.593 &  0.021           &   $R$  & DOT-3.6m \\
2458925.109   & 2020  March     16  &    18.994 &  0.019           &   $R$  & DOT-3.6m \\
2458928.209   & 2020  March     19  &    19.109 &  0.020           &   $R$  & DOT-3.6m \\
2458955.197   & 2020  April     15  &    20.673 &  0.026           &   $R$  & DOT-3.6m \\
2458956.212   & 2020  April     16  &    20.665 &  0.025           &   $R$  & DOT-3.6m \\
2458959.178   & 2020  April     19  &    20.948 &  0.027           &   $R$  & DOT-3.6m \\
\hline			  
2458898.229   & 2020  February  18  &    18.327 &  0.071           &   $I$  & ST-1.04m  \\
2458900.117   & 2020  February  20  &    18.415 &  0.025           &   $I$  & HCT-2.0m \\
2458901.133   & 2020  February  21  &    18.433 &  0.035           &   $I$  & HCT-2.0m \\
2458911.343   & 2020  March     02  &    18.574 &  0.023           &   $I$  & HCT-2.0m \\
2458912.094   & 2020  March     03  &    18.636 &  0.027           &   $I$  & HCT-2.0m \\
2458912.193   & 2020  March     03  &    18.568 &  0.022           &   $I$  & DOT-3.6m \\
2458925.109   & 2020  March     16  &    18.847 &  0.021           &   $I$  & DOT-3.6m \\
2458928.209   & 2020  March     19  &    18.929 &  0.023           &   $I$  & DOT-3.6m \\
2458955.197   & 2020  April     15  &    19.760 &  0.026           &   $I$  & DOT-3.6m \\
2458956.212   & 2020  April     16  &    20.670 &  0.067           &   $I$  & DOT-3.6m \\
2458959.178   & 2020  April     19  &    20.297 &  0.059           &   $I$  & DOT-3.6m \\
      \hline 
    \end{tabular}
  \end{center}}
\end{table}

\begin{table}
\small{
 \begin{center}
    \caption{Estimated values of $K$-corrections (in magnitudes) for different passbands are listed. The values were determined using the near-peak spectrum (at $-0.74$ d) of SN~2020ank using the {\tt SNAKELOOP} code \citep{Inserra2018b}.}
    \label{tab:k_corr}
    \addtolength{\tabcolsep}{11pt}
    \begin{tabular}{ccc}
\hline
Observed band & Rest-frame band & $K$-correction \\
     $ $  &  & (mag)  \\         
    \hline
   \hline
$U$ (Vega)  &   $UVW1$ (Vega)   &  0.57   $\pm$ 0.02   \\
$B$ (Vega)  &   $U$    (Vega)   &  0.75   $\pm$ 0.01   \\  
$g$ (AB)    &   $u$    (AB)     & $-$0.17 $\pm$ 0.01   \\
$V$ (Vega)  &   $B$    (Vega)   & $-$0.28 $\pm$ 0.01   \\
$r$ (AB)    &   $g$    (AB)     & $-$0.22 $\pm$ 0.01   \\
$R$ (Vega)  &   $V$    (Vega)   & $-$0.44 $\pm$ 0.01   \\
$I$ (Vega)  &   $R$    (Vega)   & $-$0.49 $\pm$ 0.03   \\
\hline 
    \end{tabular}
  \end{center}}
\end{table}

\subsection{The rest-frame M\textsubscript{g} light curves}
\label{sec:Mg_rest}
We compare the rest-frame $g$-band absolute magnitudes of SN~2020ank with well-studied bright (M$_{g,peak}$ $>-$20.5 mag) SLSNe~I taken from \citealt{Nicholl2015a, Nicholl2016a, DeCia2018, Kangas2017, Bose2018, Kumar2020}, and references therein (see Fig.~\ref{fig:Mg}). We also compare the $M_g$ light-curve of SN~2020ank with the only known SLSN~2011kl associated with a long-duration GRB 111209A \citep{Greiner2015,Kann2019}. For SLSNe~I having Bessell $B$-band data, the transformation equations and uncertainties by \citet{Jordi2006} have been used to obtain the SDSS $g$-band magnitudes. Fig.~\ref{fig:Mg} shows that SN~2020ank is a bright SLSN with M$_{g,peak}$ $\sim-21.84$ $\pm$ $0.10$ mag, which is closer to SN~2010kd \citep{Kumar2020} and Gaia16apd \citep{Kangas2017,Nicholl2017} within errors, whereas fainter than LSQ14bdq \citep{Nicholl2015b} and SN~2015bn \citep{Nicholl2016a}. SN~2020ank appears to have steeper pre-peak rising and post-peak decaying rates similar to that observed for other fast-evolving SLSNe~I (e.g., SN~2011ke). Further, the post-peak decay rate of SN~2020ank is also steeper in comparison to slow-evolving SLSNe~I such as SN~2010kd \citep{Kumar2020}, PTF12dam \citep{Nicholl2013}, and SN~2015bn \citep{Nicholl2016a}.

To constrain the peak brightness, rise and decay times of a larger sample of such SLSNe~I (present case and those discussed in \citealp{Nicholl2015a, DeCia2018}), we independently estimate the $M_{g,peak}$ and time taken to rise/decay by 1 mag to/from the peak absolute magnitudes ($t^{\Delta 1mag}_{rise}$/$t^{\Delta 1mag}_{fall}$) using the rest-frame $M_g$ light curves. To estimate the values of $M_{g,peak}$, $t^{\Delta 1mag}_{rise}$ and $t^{\Delta 1mag}_{fall}$, we fitted a low-order spline function to the rest-frame $M_g$ light curves (see Fig.~\ref{fig:Mg_spline}, also discussed by \citealt{DeCia2018}). SLSNe~I having less pre- or post-peak data are omitted, but a sparse extrapolation was done wherever necessary. We plot the $t^{\Delta 1mag}_{fall}$ with respect to $t^{\Delta 1mag}_{rise}$ and their sum with $M_{g,peak}$ in the left-hand and right-hand panels of Fig.~\ref{fig:trise_fall_Mg}, respectively. Generally, SLSNe~I with higher values of $t^{\Delta 1mag}_{rise}$ also display the higher values of $t^{\Delta 1mag}_{fall}$ as also discussed by \citealt{Nicholl2015a,DeCia2018}, whereas no such correlation appears between $t^{\Delta 1mag}_{rise}$ + $t^{\Delta 1mag}_{fall}$ and $M_{g,peak}$.

SN~2020ank exhibits the steepest pre-peak rising rate (lower value of $t^{\Delta 1mag}_{rise}$) in comparison to other SLSNe-I in the sample, except for the GRB-associated SLSN~2011kl \citep{Kann2019}. The right-hand panel of Fig.~\ref{fig:trise_fall_Mg} also shows that SN~2020ank exhibits a faster photometric evolution (lower value of $t^{\Delta 1mag}_{rise}$ + $t^{\Delta 1mag}_{fall}$) closer to those seen for PS1-10bzj, LSQ14mo, etc., but has the brightest peak in comparison to all other plotted fast-evolving SLSNe~I. In all, SN~2020ank is a fast-evolving SLSN with high peak brightness in the $g$-band. 

\begin{figure*}
\includegraphics[angle=0,scale=0.95]{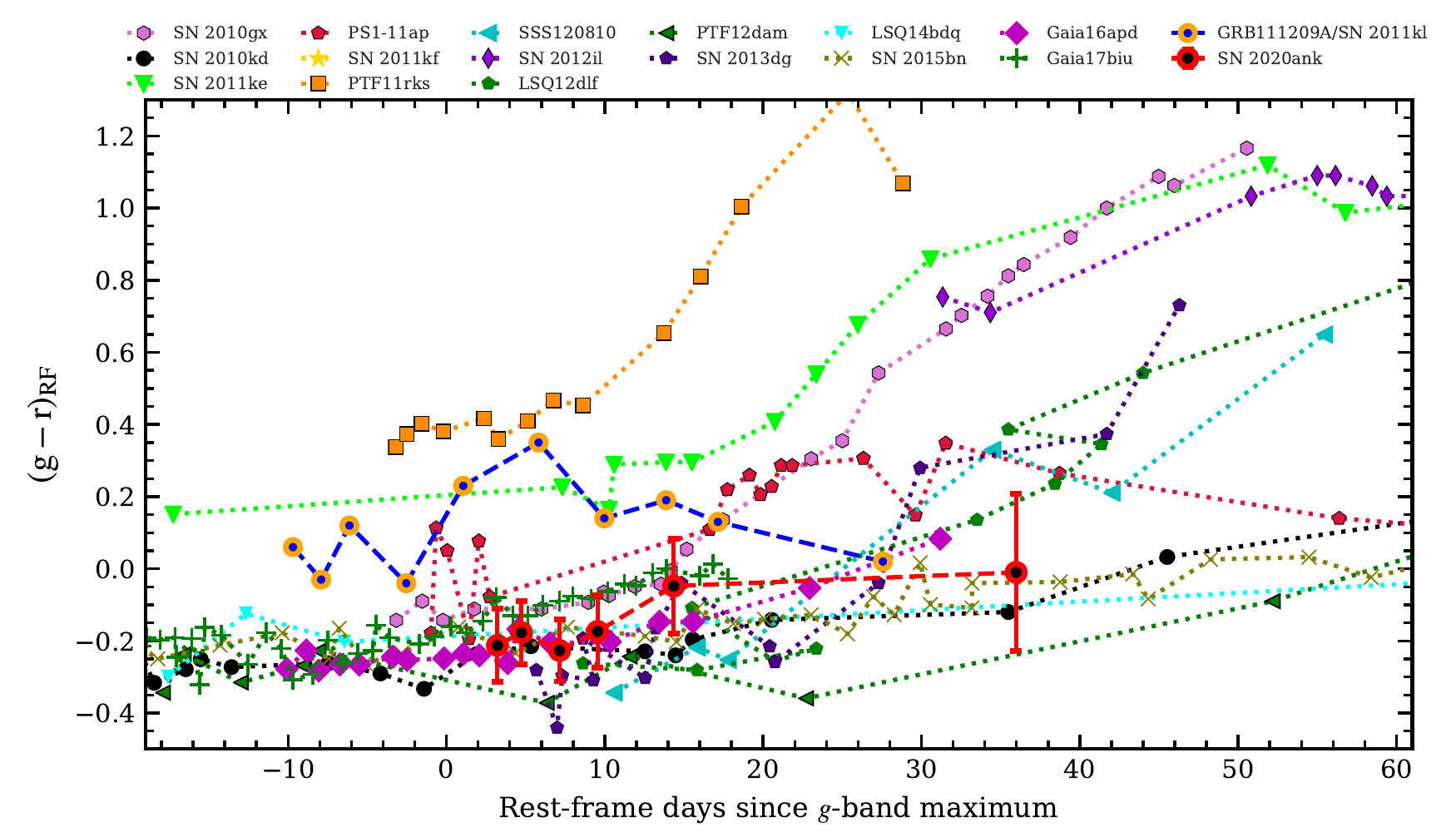}
\caption{The rest-frame  $g-r$ colour evolution of SN~2020ank (in red) along with some other well-studied SLSNe~I \citep[taken from][]{Nicholl2015a, Nicholl2016a, Kangas2017, Bose2018, DeCia2018, Kann2019, Kumar2020} is presented (Galactic extinctions and $K$-corrected). The colour evolution of SN~2020ank seems similar to those found for the slow-evolving SLSNe~I.}
\label{fig:gr}
\end{figure*}

\begin{figure}
\includegraphics[angle=0,scale=0.85]{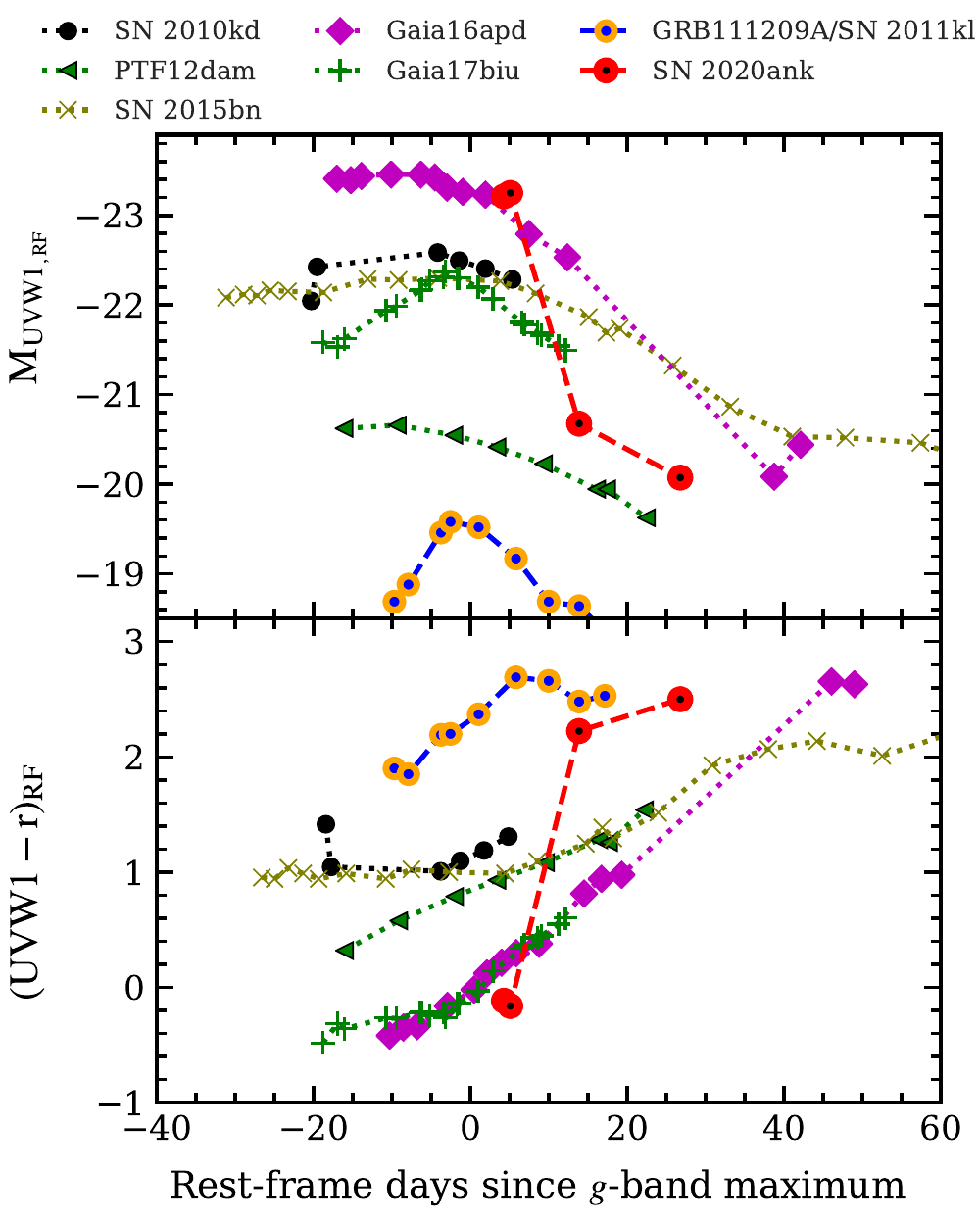}
\caption{$Upper$: The rest-frame $M_{UVW1}$ light curve of SN~2020ank (after applying Galactic extinctions and $K$-corrections) is presented and compared with the following well-studied SLSNe~I: SN~2010kd \citep{Kumar2020}, SN~2011kl \citep{Greiner2015, Kann2019}, PTF12dam \citep{Nicholl2013}, SN~2015bn \citep{Nicholl2016a}, Gaia16apd \citep{Kangas2017, Nicholl2017} and Gaia17biu \citep{Bose2018}. Near the peak, $UVW1$ brightness of SN~2020ank seems even higher to that observed in the case of Gaia16apd. $Lower$: The rest-frame $UVW1-r$ colour evolution of SN~2020ank is compared with the same sample of SLSNe~I discussed above in the AB magnitude system. Near the peak, SN~2020ank exhibits bluer colour in comparison to other SLSNe~I and turns redder very sharply with time.}
\label{fig:UV_bright}
\end{figure}

\subsection{The rest-frame g - r colour evolution}\label{sec:g-r}

During the photospheric phase, optical-NIR colours are useful probes to understand the temperature evolution of SLSNe. Due to the unavailability of the rest-frame $r$-band data for SN~2020ank, the rest-frame (Galactic extinction and $K$-corrected) $V$ and $R$ bands data were transformed to the $r$-band magnitudes using the transformation equation and uncertainties given by \citet{Jordi2006}. So, here we compare the rest-frame $g-r$ colour evolution of SN~2020ank with that observed for the well-studied SLSNe~I (see Fig.~\ref{fig:gr}). The slow-evolving SLSNe~I (SN~2010kd, PTF12dam, SN~2015bn, etc.) appear to have shallower rising (from blue to red) colour evolution (from $\sim$--0.4 to 0.2 mag in a time range of $\sim$--20 to +100 d). In contrast, the fast-evolving SLSNe~I (e.g., SN~2010gx, SN~2011ke, 2012il, etc.) show $g-r$ colour ranging from $\sim$\,0.0 to $\gtrsim$\,1.2 mag in the same temporal bin. SN~2020ank presents a bluer $g-r$ colour evolution from $\sim$+3 to +36 d spanning the range from $-$0.3 to 0.1 mag. Overall, the $g-r$ colour of the fast-evolving SN~2020ank is closer to the slow-evolving SLSNe~I. The rest-frame $UVW1-r$ colour evolution of SN~2020ank is also discussed in Section~\ref{sec:uv_bright}.

\subsection{Comparison of UV-brightness}\label{sec:uv_bright}

In this section, we compare the rest-frame equivalent $UVW1$ brightness and evolution of the $UVW1 - r$ colour of SN~2020ank with other well-studied SLSNe~I (see Fig.~\ref{fig:UV_bright}). Near the peak, $UVW1$ brightness of SN~2020ank appears to be comparable to Gaia16apd, the most $UV$ bright SLSN to date \citep{Nicholl2017} except the most luminous SLSN ASASSN-15lh \citep{Dong2016}. With time, the $UVW1$ flux of SN~2020ank decays very sharply and becomes fainter in comparison to other presented SLSNe~I (except PTF12dam and SN~2011kl), see the upper panel of Fig.~\ref{fig:UV_bright}. One of the possible reasons behind this excess UV flux of SN~2020ank near the peak may be the lower production of heavier group elements during the explosion, as suggested by \citet{Yan2017a} in the case of Gaia16apd (see also \citealp{Mazzali2016}). The other possible reason could be a short-lived powering source adding extra luminosity component towards the $UV$ or lower natal metallicity. However, these plausible reasons for explaining the excess $UV$ flux were not found suitable in the case of Gaia16apd \citep{Nicholl2017}. This is because of its similar metallicity and degree of spectral absorption features to the other SLSNe~I having a lower $UV$ flux. In the case of Gaia16apd, the most likely possible reason might be a central engine as a power source (it could be a spin-down millisecond magnetar with lower spin period and comparatively low mass or a mass accreting BH; \citealt{MacFadyen1999}), which could explain both the overall luminosity and the excess $UV$ flux \citep{Nicholl2017}. For SN~2020ank as well, the central engine based powering source as a possible mechanism for the observed $UV$ excess near the peak is well in agreement with semi-analytical light-curve modelling results discussed later in Section~\ref{sec:MINIM}.

In the lower panel of Fig.~\ref{fig:UV_bright}, we compare the $UVW1 - r$ colour evolution of SN~2020ank with the well-studied SLSNe~I. The rest-frame $UVW1 - r$ colour curves of PTF12dam, SN~2015bn, and Gaia16apd are taken from \citet{Nicholl2017}, whereas calculated independently for SN~2010kd and Gaia17biu. Due to the unavailability of SDSS $r$-band data for SN~2020ank and SN~2010kd, we obtained the $r$-band data from the available $V$ and $R$ bands data using the transformation equations and uncertainties given by \citet{Jordi2006}. Before estimating colours, the $UVW1$ magnitudes were converted from Vega to AB system using the zero points adopted from \cite{Breeveld2011}. Near the peak, $UVW1 - r$ colour of SN~2020ank is closer to those observed in the case of Gaia16apd and Gaia17biu; however, it is $\sim$1$-$2.5 mag bluer in comparison to other presented SLSNe~I. With time, $UVW1 - r$ colours of SN~2020ank turn redder quite sharply, indicating a rapid drop in temperature, also in agreement with the estimate of temperature using the BB fitting to the photometric spectral energy distribution (SED), see Section~\ref{sec:bolo}.

\begin{figure*}
\includegraphics[angle=0,scale=1.0]{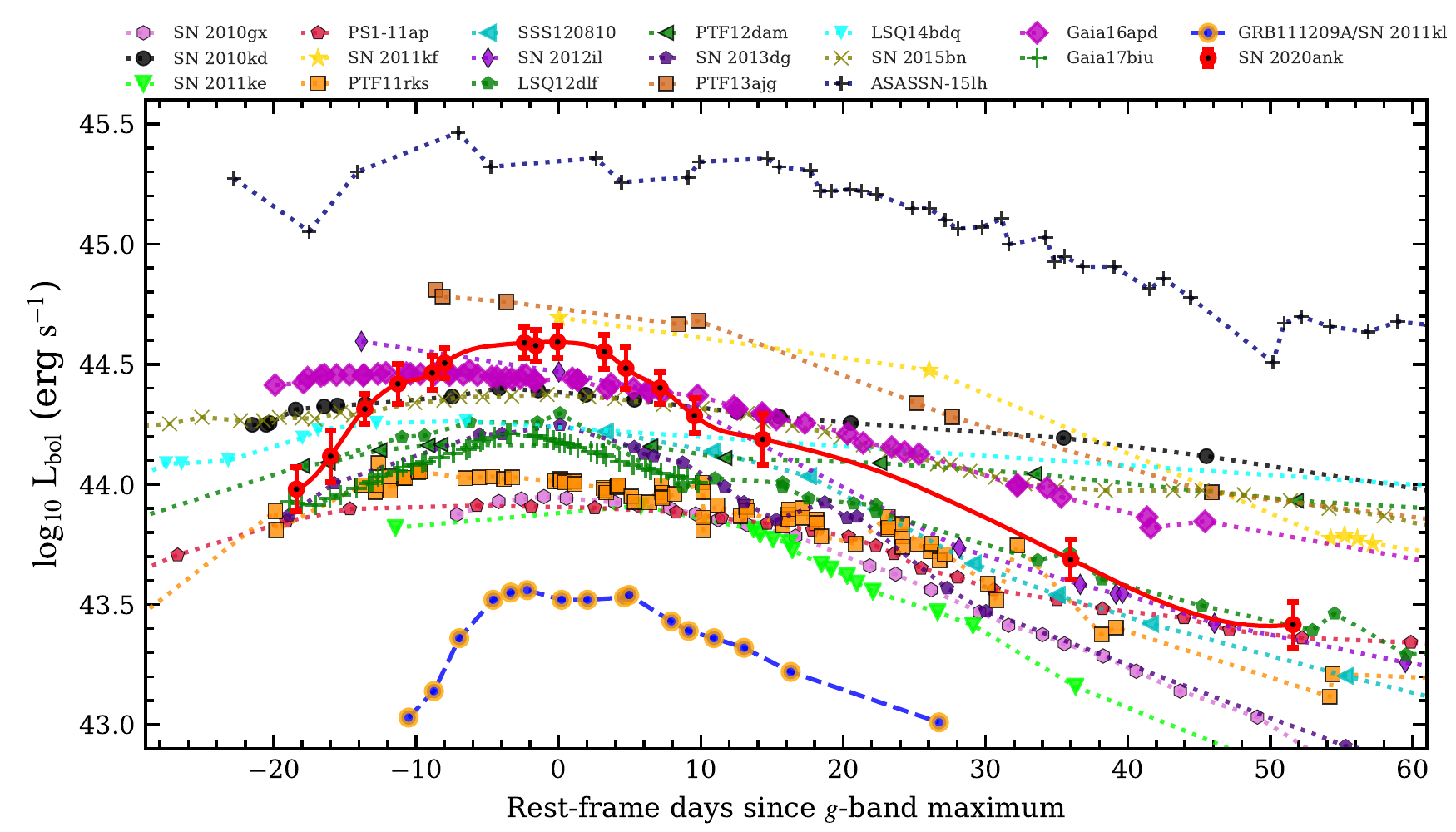}
\caption{The rest-frame bolometric light curve of SN~2020ank ($UV$ to $NIR$) and its comparison with a sample of other well-studied SLSNe~I taken from \citealt{Vreeswijk2014, Nicholl2015a, Dong2016, Nicholl2016a, Kangas2017, Bose2018, DeCia2018, Kann2019, Kumar2020} are presented. The comparison indicates that SN~2020ank is one of the brightest SLSN though fainter than SN~2011kf \citep{Inserra2013}, PTF13ajg \citep{Vreeswijk2014}, and ASASSN-15lh \citep{Dong2016}.}
\label{fig:full_bolo}
\end{figure*}

\begin{figure}
\includegraphics[angle=0,scale=0.7]{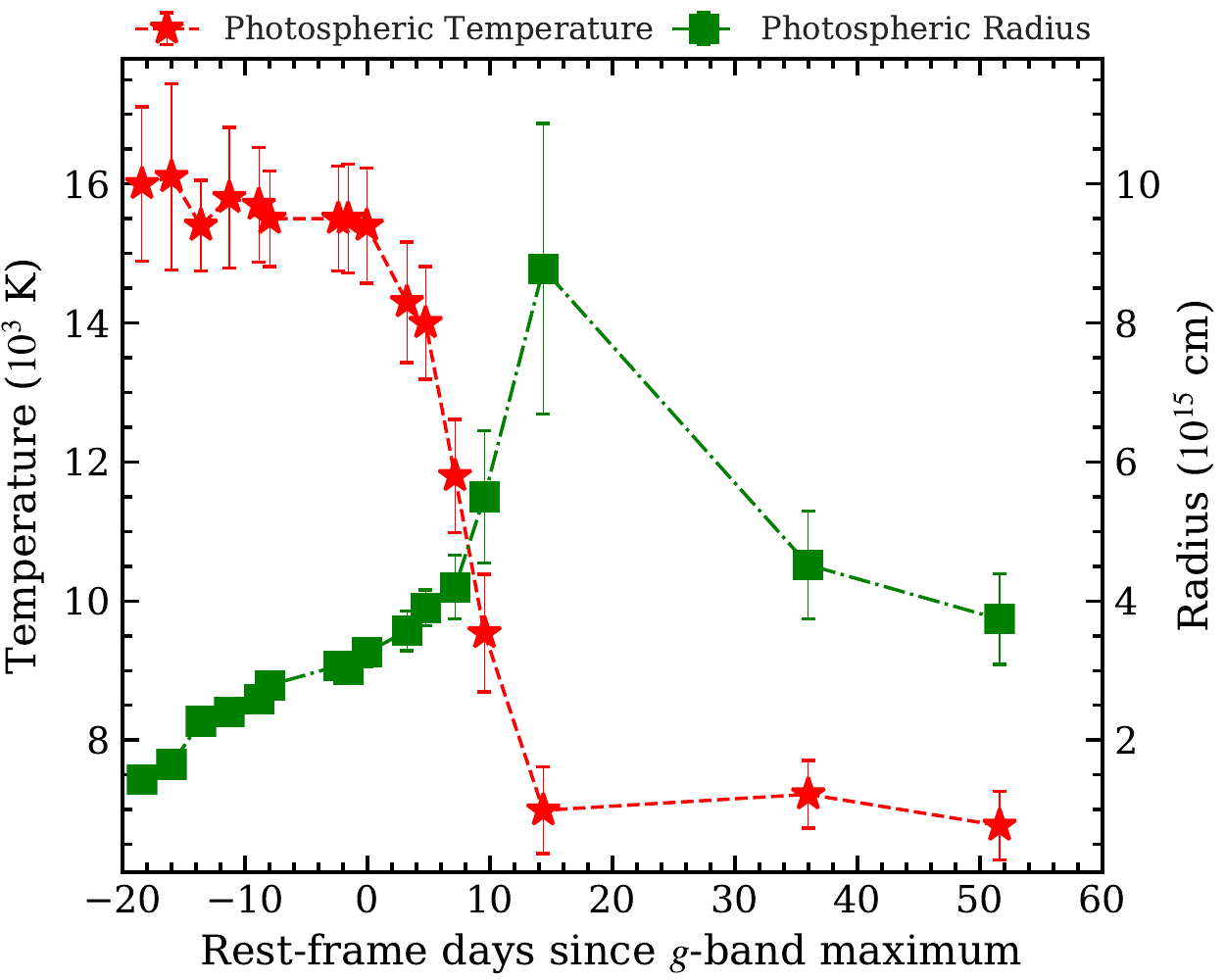}
\caption{The photospheric temperature ($T_{BB}$) and radius ($R_{BB}$) evolution of SN~2020ank are presented, through BB fit to the photometric data using the {\tt Superbol} code \citep{Nicholl2018}.}
\label{fig:boltemprad}
\end{figure}

\begin{figure*}
\includegraphics[angle=0,scale=1.1]{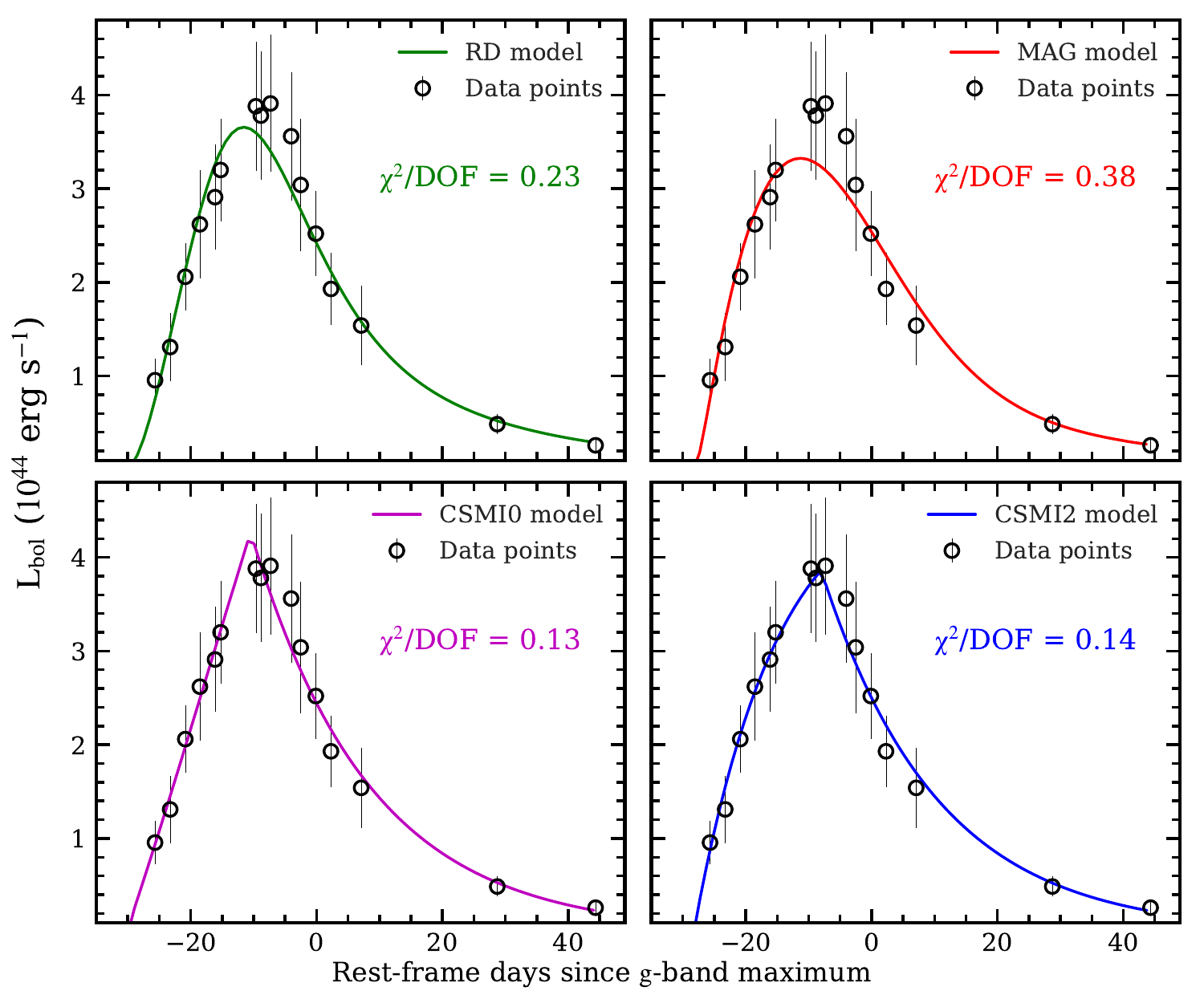}
\caption{Semi-analytical light-curve models (RD, MAG, CSMI0, and CSMI2) are fitted to the bolometric light curve of SN~2020ank using the {\tt MINIM} code \citep{Chatzopoulos2013}. For all the model fittings, the $\kappa$ = 0.1 cm$^2$ g$^{-1}$ is adopted. All four discussed models well reproduced the bolometric light curve of SN~2020ank with $\chi^2$/DOF $\sim$1; but, the physical reliability of the derived parameters suggested the MAG model as the most suitable one for SN~2020ank.}
\label{fig:minim}
\end{figure*}

\section{Bolometric light curve} \label{sec:bolo}

We generate the rest-frame quasi-bolometric ($UVW1, u, U, B, g, V,$ and $R$ bands) light curve of SN~2020ank using a Python-based code {\tt Superbol} \citep{Nicholl2018}. To compute the multiband fluxes, interpolation or extrapolation was done wherever necessary assuming constant colours to get the magnitudes at the individual epochs using standard methods. The uncertainties in the calculation of zero points are also taken care of by adding 3\% uncertainties to the bolometric luminosity error as is suggested by the software release of {\tt Superbol}\footnote{\url{https://github.com/mnicholl/superbol}}. The pseudo-bolometric ($UVW1$ to $R$) light curve of SN~2020ank presents a peak luminosity ($L_{max}$) of $\sim$(1.89 $\pm$ 0.14) $\times$ 10$^{44}$ erg s$^{-1}$. To include the expected flux contribution from $UV$ and $NIR$ regions, we extrapolated the SED (here the BB model) by integrating over the observed fluxes and obtained the full bolometric light curve. The full bolometric light-curve ($UV$ to $NIR$) of SN~2020ank (in red) between $\sim -$20 and +50 d, derived so, exhibited $L_{max}$ of $\sim$(3.89 $\pm$ 0.69) $\times$ 10$^{44}$ erg s$^{-1}$ at MJD $\approx~58892.8$, overall consistent with other well-studied SLSNe~I, see Fig.~\ref{fig:full_bolo}. In the case of SN~2020ank, the bolometric light curve is nearly symmetric around the peak with the $L^{rise}_{max}$/e $\approx$ $L^{fall}_{max}$/e $\approx$ 15 d (time taken to rise/decay by $L_{max}$/e to/from the $L_{max}$).

For an extensive comparison of bolometric luminosities, the sample used in Fig.~\ref{fig:Mg} was supplemented with data of PTF13ajg \citep{Vreeswijk2014} and ASASSN-15lh \citep{Dong2016} along with the independently generated bolometric light curves of PTF11rks, SN~2011kf and SN~2012il \citep{Inserra2013}. The comparison shows that SN~2020ank is one of the brightest SLSN with a peak bolometric luminosity higher in comparison to SN~2010kd, PTf12dam, SN~2015bn, etc., but fainter than SN~2011kf, PTF13ajg, and the most luminous SLSN ASASSN-15lh \citep{Dong2016} in the sample. As obtained from the rest-frame $M_g$ light-curves comparison, SN~2020ank appears to have high pre-peak rising and post-peak decay rates of its bolometric light curve, similar to that of fast-evolving SLSNe~I (e.g., SN~2010gx and SN~2011ke). In all, SN~2020ank is a bright SLSN having a bell-shaped light curve around the peak with a fast-evolving behaviour.

Fig.~\ref{fig:boltemprad} shows the evolution of BB temperature (T$_{BB}$) and radius ($R_{BB}$) of SN~2020ank. The $T_{BB}$ and $R_{BB}$ values are calculated by modelling the photometric SED at individual epochs by fitting a BB function using the {\tt Superbol} code \citep{Nicholl2018}. From $\sim -20$ d to peak, the $T_{BB}$ of SN~2020ank seems to be constant around $\sim$16000 K, whereas from peak to $\sim$+15 d it sharply decays with a rate of $\sim$600 K per day. At later epochs (after $\sim$+15 d), the $T_{BB}$ appears to be constant again at $\sim$7000 K. The near peak T$_{BB}$ of SN~2020ank is higher in comparison to well-studied slow-evolving SLSNe~I PTF12dam and SN~2015bn \citep{Chen2015,Nicholl2016a}, nearly consistent with intermediate decaying Gaia16apd \citep{Kangas2017,Nicholl2017}, and lower than the most luminous SLSN ASASSN-15lh \citep{Dong2016}. On the other hand, from $\sim -$20 to +15 d the value of $R_{BB}$ for SN~2020ank increases from $\sim$1.3 $\times$ $10^{15}$ to 8.8 $\times$ $10^{15}$ cm, thereafter, up to $\sim$+50 d, it decreases to $\sim$3.4 $\times$ $10^{15}$ cm.

\subsection{Light-curve modelling using {\tt MINIM}} \label{sec:MINIM}

We attempt to reproduce the bolometric light curve of SN~2020ank with the RD, MAG, constant density CSMI (CSMI0), and wind-like CSMI (CSMI2) semi-analytical light-curve models (see Fig.~\ref{fig:minim}) using the {\tt MINIM} code \citep{Chatzopoulos2013}. {\tt MINIM} is a general-purpose fitting code that finds the global solution for non-linear $\chi^2$ fitting. It uses the Price algorithm \citep{Brachetti1997}, a controlled random-search technique, to look for the global minimum of the $\chi^2$ hyper-surface within the allowed parameter volume. After that, parameters generated by the Price algorithm for the lowest $\chi^2$ value are fine-tuned by the Levenberg-Marquardt algorithm \citep{Jorge1978}, which gives the final set of parameters for the best-fitted model. The uncertainties in the parameters are estimated using the standard deviation of the random vectors around the global minimum. Details about the above discussed models, {\tt MINIM} code and fitting procedures are described in \citet{Chatzopoulos2012, Chatzopoulos2013}.

For all the models discussed above, we adopted the electron-scattering opacity, $\kappa$ = 0.1 cm$^2$ g$^{-1}$, by considering half ionized elements for SLSNe~I \citep{Inserra2013, Andrea2018}. For RD and MAG models, the $M_{ej}$ values are estimated using equation 10 of \citet{Chatzopoulos2012}, where the integration constant ($\beta$) was considered to be 13.8. In the case of RD model, $v_{\rm exp}$ is taken equals to 12000 km s$^{-1}$ as obtained from the spectral analysis of SN~2020ank discussed in Section~\ref{sec:synapps}. However, for the MAG model, $v_{\rm exp}$ was given by the {\tt MINIM} itself as a fitting parameter. The initial period of the new-born magnetar in ms ($P_i$) is given by $P_i$ = 10 $\times$ $(\frac{2 ~ \times ~ 10^{50} ~ erg ~ s^{-1}}{E_p})^{0.5}$, where $E_p$ is the magnetar rotational energy in erg. The magnetic field of the millisecond magnetar in Gauss units ($B$) is estimated by $B$ = 10$^{14}$ $\times$ $(\frac{130 ~ \times ~ P_i^2}{t_{p,yr}})^{0.5}$, where $t_{p,yr}$ is the magnetar spin-down time-scale in years.

\begin{table*}
% \small{
 \begin{center}
  \begin{threeparttable}
    \caption{Best-fitting parameters for the RD, MAG, CSMI0, and CSMI2 models obtained using the {\tt MINIM} fitting.}
    \label{tab:minim}
    \addtolength{\tabcolsep}{2pt}
    \begin{tabular}{c c c c c c c c c c}

    \hline \hline
          &  &  &  & \textbf{RD model}& & & & & \\
      $A_{\gamma}$ $^a$ & $M_{\rm ej}$ $^b$ & & $t_d$ $^c$ &  $M_{\rm Ni}$ & & & & $\chi^2$/DOF &\\
        &($M_\odot$) & & (d) & ($M_\odot$) & & & & & \\
      \\
     9.20 $\pm$ 1.34 & 2.62 $\pm$ 0.38 & & 21.66 $\pm$ 1.55 &  26.71  $\pm$ 2.55 &  & & & 0.23 & \\

    \hline \hline
    
              &  &  &  & \textbf{MAG model}& & & & & \\
     $R_0$ & $M_{\rm ej}$ & $E_p$ $^d$ & $t_d$ & $t_p$ $^e$ & $v_{\rm exp}$ &  $P_i$ & $B$ & $\chi^2$/DOF & \\

      ($10^{13}$ cm) & ($M_\odot$) & ($10^{51}$ erg) & (d) & (d) & ($10^3$ km s$^{-1}$) &  (ms) & ($10^{14}$ G) & &\\
      \\
       \textbf{0.28} $\pm$ \textbf{0.22} & \textbf{3.58} $\pm$ \textbf{0.04} & \textbf{4.02}  $\pm$ \textbf{0.19} & \textbf{25.01} $\pm$ \textbf{0.16}  & \textbf{2.79} $\pm$ \textbf{0.35} & \textbf{12.27} $\pm$ \textbf{0.91} & \textbf{2.23}  $\pm$ \textbf{0.51}  & \textbf{2.91} $\pm$ \textbf{0.07} & \textbf{0.38}  &  \\

    \hline \hline

    $R_p$ $^f$ & $M_{\rm ej}$ & $M_{\rm csm}$ $^g$ & $\dot{M}$ $^h$ &    & $v_{\rm exp}$ &  & & $\chi^2$/DOF &\\
     ($10^{13}$ cm) & ($M_\odot$) & ($M_\odot$) & ($M_\odot$ yr$^{-1}$) &  & ($10^3$ km s$^{-1}$) & & &\\
      \\
              &  &  &  & \textbf{CSMI0 model}& & & & & \\
       263.20  $\pm$  45.57 & 46.13  $\pm$ 7.78 & 9.13 $\pm$ 1.56 & 54.95 $\pm$ 11.32 &   & 29.69 $\pm$ 1.84 &  & & 0.13 &\\
              &  &  &  & \textbf{CSMI2 model}& & & & & \\    
       312.90 $\pm$ 19.87 & 45.11 $\pm$ 8.93 & 13.12 $\pm$ 0.83 & 0.58 $\pm$ 0.05 &   & 23.23 $\pm$ 0.59 &  &  & 0.14 &\\
    \\
      
      \hline 
    \end{tabular}
    \begin{tablenotes}[para,flushleft]
        $^a$ $A_\gamma$: optical depth for the gamma-rays measured after the 10 d of explosion.
        $^b$ $M_{\rm ej}$: ejecta mass (in $M_\odot$).
        $^c$ $t_d$: effective diffusion-timescale (in d).    
        $^d$ $E_p$: magnetar rotational energy (in $10^{51}$ erg).
        $^e$ $t_p$: magnetar spin-down timescale (in d).
        $^f$ $R_p$: progenitor radius before the explosion (in $10^{13}$ cm).
        $^g$ $M_{\rm csm}$: CSM mass (in $M_\odot$).
        $^h$ $\dot{M}$: progenitor mass-loss rate (in $M_\odot$ yr$^{-1}$).
    
    \end{tablenotes}
  \end{threeparttable}
  \end{center}
\end{table*}

\begin{figure*}
\includegraphics[angle=0,scale=0.90]{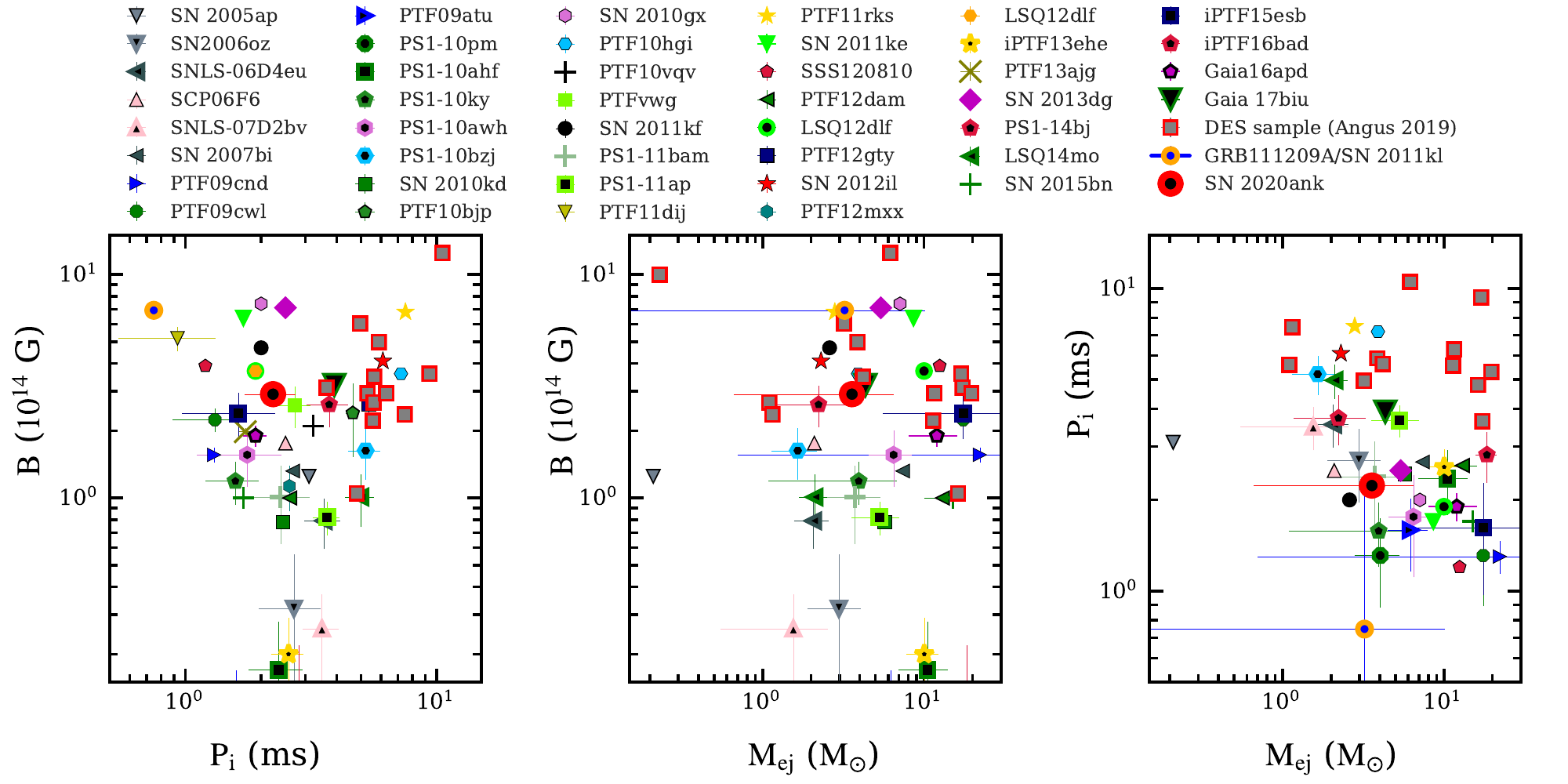}
\caption{Scatter plot showing the comparison of $P_i$, $B$, and M$_{ej}$ values derived for SN~2020ank using the semi-analytical light-curve {\tt MINIM} modelling with the sample of the well-studied SLSNe~I is presented.}
\label{fig:Mej_Ek}
\end{figure*}

It is evident from Fig.~\ref{fig:minim} that all four models (RD, MAG, CSMI0, and CSMI2) fit the data adequately, i.e., with $\chi^2$/DOF $\sim$1. Here, we caution that in the present case, the $\chi^2$/DOF is used as an indicator for selecting the model parameters that fit the data best and not as a statistical probe for judging the significance of the models. In this situation, the best $\chi^2$ value solely cannot be used as a criterion to declare the most probable model. Therefore, we look for the feasibility of various physical parameters retrieved from different models. The parameters obtained from the {\tt MINIM} modelling using the four discussed models are tabulated in Table~\ref{tab:minim}. In the RD model, the calculated $M_{ej}$ is lower than $M_{Ni}$ (see Table~\ref{tab:minim}), an unphysical case, so we exclude RD's possibility as a possible powering mechanism for SN~2020ank. The CSMI0 and CSMI2 models also well reproduced the bolometric light curve of SN~2020ank though some of the derived parameters seem quite unphysical (e.g., very high values of $R_p$, $M_{ej}$, $v_{exp}$, and $\dot{M}$). For both the models, $M_{\rm ej}$ and $v_{\rm exp}$ values are very high, which correspond to the kinetic energy of a few $\times$ 10$^{53}$ erg, an inadmissible amount for CCSNe \citep{Janka2012}. Besides this, values of $M_{\rm ej}$ and $v_{\rm exp}$ obtained from CSMI0 and CSMI2 model fits are also higher than those from our spectral analysis in Section~\ref{sec:spectra}. The estimated values of $R_{p}$ are also of the order of $\sim$10$^{15}$ cm ($\gtrsim$35000 $R_{\odot}$; see Table~\ref{tab:minim}), which is very high and close to the radius of the H-rich CSM shells, making it unphysical in the present case \citep{Yan2017b,Nicholl2019}. The CSMI0 model gives rise to an unacceptable amount of $\dot{M}$ $\sim$55 $M_\odot$ yr$^{-1}$. Though the CSMI2 model has a comparatively lower $\dot{M}$ ($\sim$0.6 $M_\odot$ yr$^{-1}$), but this is also an unreasonably higher rate before the explosion. All these factors indicate that CSMI0 and CSMI2 models are not favoured to explain the light curve of SN~2020ank. On the other hand, around the peak the MAG model poorly fits the bolometric light curve of SN~2020ank because of its bell-shaped ($L^{rise}_{max}$/e $\approx$ $L^{fall}_{max}$/e $\approx$ 15 d) layout \citep{Chatzopoulos2019}. However, the MAG model presents a reasonable set of physical parameters among all these discussed four {\tt MINIM} models. The estimated parameters by the MAG model for SN~2020ank (see Table~\ref{tab:minim}) are within the range of parameters found for other well-studied SLSNe~I (see Section~\ref{sec:comp_Pi_B_Mej}) and close with the corresponding parameters obtained from the spectral analysis in Section~\ref{sec:spectra}. The spin-down millisecond magnetar as a possible powering source for SN~2020ank is also consistent with the excess $UV$ flux of SN~2020ank near the peak favouring a central engine based power source (see Section~\ref{sec:uv_bright}).

In summary, based on our fitting we consider the MAG model as the most probable one because (1) it fits the data well ($\chi^2$/DOF $\sim$1), and (2) its parameters are realistic and closer to the ones inferred from the spectral modelling ($M_{\rm ej}$, $v_{\rm exp}$; see Section~\ref{fig:spectra}). So, the spin-down millisecond magnetar is found to be the most suitable powering source for SN~2020ank with $P_i$ of $\sim$2.23 $\pm$ 0.51 ms and $B$ $\sim$(2.91 $\pm$ 0.07) $\times$ $10^{14}$ G, giving rise to an ejected mass of $\sim$3.58 $\pm$ 0.04 $M_{\odot}$. The MAG model suggests the value of progenitor radius ($R_0$) $\sim$(2.8 $\pm$ 2.2) $\times$ 10$^{12}$ cm, nearly forty times the solar radius, whereas other parameters are tabulated in Table~\ref{tab:minim} in bold.

\subsubsection{Comparison of derived physical parameters with other SLSNe~I}
\label{sec:comp_Pi_B_Mej}
In this section, we compare the $P_i$, $B$, and $M_{ej}$ of SN~2020ank estimated through semi-analytical light-curve modelling using {\tt MINIM} with those found in case of other well-studied SLSNe~I (see Fig.~\ref{fig:Mej_Ek}): SN~2005ap, SCP06F6, SN~2007bi \citep{Chatzopoulos2013}, SN~2010gx, PTF10hgi, PTF11rks, SN~2011ke, SN~2011kf, SN~2012il \citep{Inserra2013}, PTF12dam \citep{Nicholl2013}, LSQ12dlf, SSS120810, SN~2013dg \citep{Nicholl2014}, SN~2015bn \citep{Nicholl2016a}, Gaia16apd \citep{Kangas2017}, SN~2006oz, SNLS-07D2bv, SNLS-06D4eu, PTF09atu, PS1-10pm, PS1-10ahf, PS1-10ky, PS1-10awh, PS1-10bzj, PS1-11bam, PS1-11ap, LSQ12dlf, iPTF13ehe, PS1-14bj, LSQ14mo, iPTF15esb, iPTF16bad \citep{Nicholl2017a}, Gaia17biu \citep{Wheeler2017}, PTF09cnd, PTF09cwl, PTF10bjp, PTF10vqv, PTF10vwg, PTF11dij, PTF12gty, PTF12mxx, PTF13ajg \citep{DeCia2018}, SN~2010kd \citep{Kumar2020}, and the sample of SLSNe~I from the Dark Energy Survey \citep[DES;][]{Angus2019}. We also consider SN~2011kl for comparison, as this is the only case having confirmed association of an SLSN with a long GRB, favouring a central engine driven powering source \citep{Bersten2016ApJ, Kann2019, Lin2020}. However, we also caution that distinct methods were used in different studies to estimate these parameters.

In the left-hand panel of Fig.~\ref{fig:Mej_Ek}, we present $P_i$ versus $B$ derived for SN~2020ank with other well-studied SLSNe~I discussed above, whereas middle and right-hand panels show comparisons of the $M_{ej}$ versus $B$ and $P_i$, respectively. Most of the SLSNe~I presented here have $P_i$ and $B$ values varying in the range of $\sim$1$-$8 ms and $\sim$(1$-$8) $\times$ 10$^{14}$ Gauss, respectively, whereas $M_{ej}$ values vary from $\sim$1 to 25 M$_\odot$. Overall, most of the slow-evolving SLSNe~I (e.g., SN~2010kd and SN~2015bn) appear to have larger $M_{ej}$ values in comparison to those exhibited by the fast-evolving SLSNe~I  (e.g., SN~2010gx and SN~2011kf), as also stated by \cite{Reka2020b}. In the case of SN~2020ank, the values of $P_i$ ($\sim$2.23 $\pm$ 0.51 ms), $B$ ($\sim$(2.91 $\pm$ 0.07) $\times$ $10^{14}$ G), and $M_{ej}$ ($\sim$3.58 $\pm$ 0.04 $M_{\odot}$) are consistent with those found for other well-studied SLSNe~I. However, the $P_i$ value of SN~2020ank is closer to PS1-10ahf and PS1-11bam, the $B$ is similar to DES14X3taz and DES17C3gyp \citep{Angus2019}, and the $M_{ej}$ is consistent with that derived for PS1-11bam and LSQ12dlf. Whereas, the $M_{ej}$ estimated for SN~2020ank ($\sim$3.58 $\pm$ 0.04 $M_{\odot}$) is also closer to the $M_{ej}$ obtained for SN~2011kl \citep[$\sim 3.22 \pm 1.47 M_\odot;$][]{Lin2020}. The value of $P_i$ for SN~2020ank is higher whereas $B$ value is lower as compared to those obtained for SN~2011kl ($P_i \sim$0.36$-$0.78 ms and $B \sim$(3.1$-$6.8) $\times$ 10$^{14}$ G, respectively, see \citealt{Bersten2016ApJ,Lin2020}). This is evident in the left-hand panel of Fig.~\ref{fig:Mej_Ek} that the GRB associated SN~2011kl shows the highest value of $B$ and lowest value of $P_i$ among all the SLSNe~I of the sample (except for SN~2010gx and SN~2013dg).
 
\section{Spectroscopic analysis of SN~2020\lowercase{ank}} \label{sec:spectra}

\begin{figure}
\includegraphics[angle=0,scale=0.9]{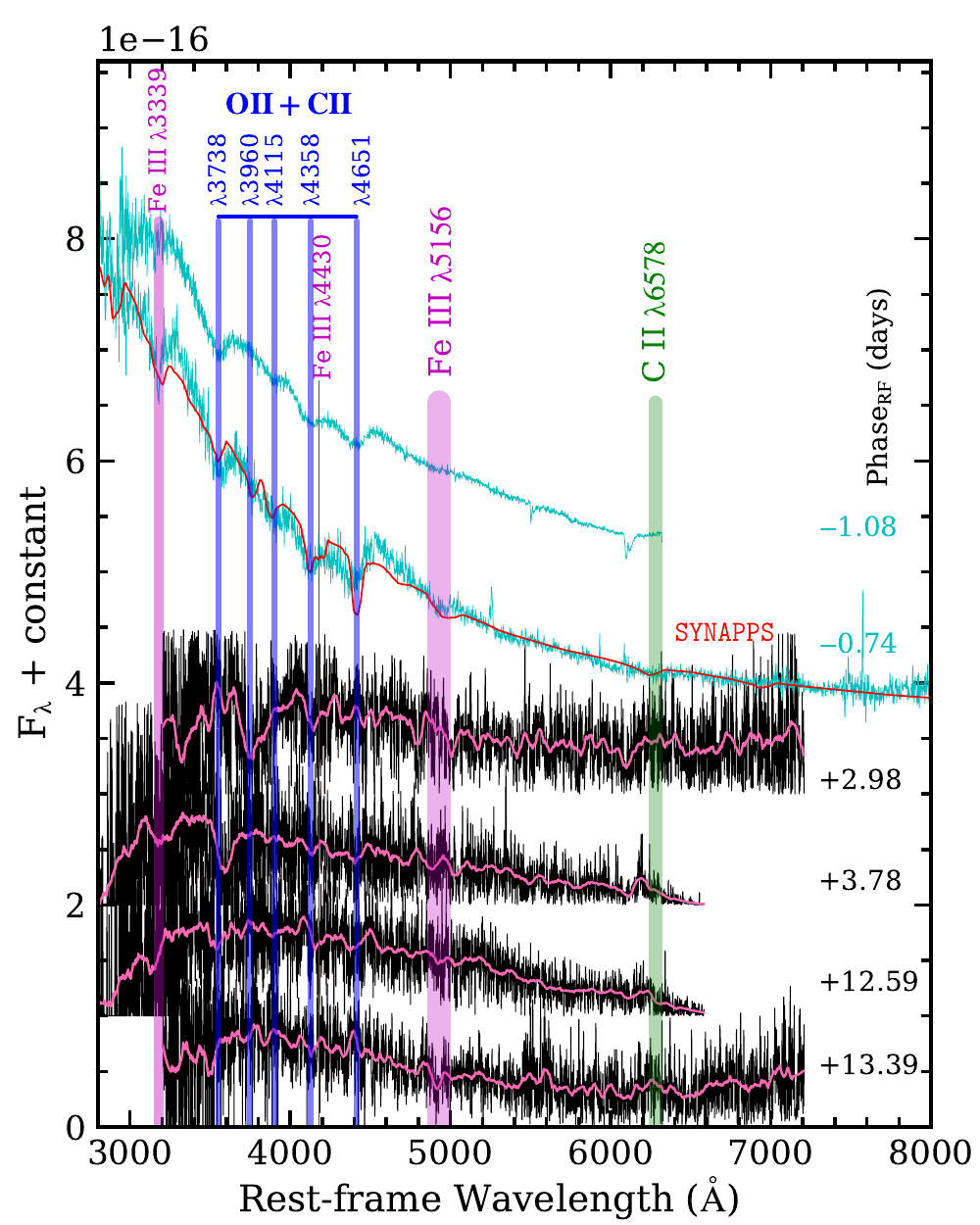}
\caption{Galactic extinction corrected spectra of SN~2020ank in the rest-frame wavelengths are shown. The phase with respect to the \textit{g}-band maximum is also mentioned on the right-hand side of each spectrum in their respective colours. The HCT-2.0m spectra are smoothed using the Savitzky–Golay smoothing method \citep{Quimby2018}, and the smoothed spectra are overplotted with the magenta colour. The near-peak spectrum (at $-$0.74 d) is regenerated using the {\tt SYNAPPS} code \citep{Thomas2011} and shown in the red. The near-peak spectra are dominated by the W-shaped O\,{\sc ii} features, whereas the weak C\,{\sc ii} and Fe\,{\sc iii} lines can also be traced.}
\label{fig:spectra}
\end{figure}

In this section, we investigate the spectral properties of SN~2020ank using spectral observations taken at six epochs (see Section~\ref{sec:reduction} and Table~\ref{tab:tablespec}). Spectra of SN~2020ank, spanning a duration of about 14 d starting from $\sim -$1 to +13 d (rest frame), are plotted in Fig.~\ref{fig:spectra}. The phase of each individual spectrum with respect to the $g$-band maximum (in d) is marked on the right-hand side of each spectrum in respective colours. All the presented spectra are shifted to the rest-frame wavelengths and corrected for the Galactic extinction value of $E(B-V)$ = 0.023 mag, whereas the host galaxy extinction was assumed negligible. The publicly available spectra obtained from the GTC-10.4m (at $-1.08$ d) and the P200-5.1m (at $-0.74$ d) are shown in cyan, however, the spectra observed from the HCT-2.0m (from +2.98 to +13.39 d) are presented in black. Unfortunately, spectra from the HCT-2.0m have a poor signal-to-noise ratio. Therefore, these spectra were smoothed using the Savitzky--Golay method by fitting the third-order polynomial function for each $\lambda$ in the range $\lambda$ $-$ $\lambda$/100 $<$ $\lambda$ $<$ $\lambda$ + $\lambda$/100 and shown with the magenta colour overplotted to the HCT-2.0m spectra (see Fig.~\ref{fig:spectra}). The P200-5.1m spectrum at $-0.74$ d is also smoothed in the same fashion in the range $\lambda$ $-$ $\lambda$/5 $<$ $\lambda$ $<$ $\lambda$ + $\lambda$/5, just for the clarity of the spectral features. Details about Savitzky--Golay smoothing technique are described in \citet{Quimby2018}. 

The near-peak spectrum (at $-$0.74 d) of SN~2020ank exhibits a bluer continuum, fitted with a BB function having temperature $\sim$14800 $\pm$ 300 K. Line identification is done following \citealt{Pastorello2010, Yan2017a, Quimby2018} and using the spectral fitting codes {\tt SYNAPPS} and {\tt SYN++} \citep{Thomas2011}. The {\tt SYNAPPS} spectral fitting is attempted for the spectrum at $-$0.74 d, and the modelled spectrum is shown with the red colour, see Fig.~\ref{fig:spectra}, details about the {\tt SYNAPPS} spectral fitting are discussed in Section~\ref{sec:synapps}. The spectral features are highlighted with the vertical bands at the observed wavelengths, and their respective rest-frame wavelengths are written on the top. The near-peak spectra (at $-$1.08 and $-$0.74 d) of SN~2020ank are dominated by the W-shaped O\,{\sc ii} features ($\lambda$3737.59, $\lambda$3959.83, $\lambda$4115.17, $\lambda$4357.97 and $\lambda$4650.71) discussed by \citet{Quimby2018} and exhibiting a clear absence of H and He lines. However, weak signatures of C\,{\sc ii} ($\lambda$3921, $\lambda$4267, $\lambda$6578 and $\lambda$7234) and Fe\,{\sc iii} ($\lambda\lambda$ 3339, 4430 and $\lambda$5156) are also traced. In the near-peak spectra, the O\,{\sc ii} absorption features are highly blueshifted with respect to their rest-frame wavelengths, indicating an equivalent velocity of $\sim$17500 km s$^{-1}$, and C\,{\sc ii} ($\lambda$6578) observed at $\sim$6220 \AA~ with a velocity around 16500 km s$^{-1}$. On the other hand, Fe\,{\sc iii} ($\lambda$5156) observed at $\sim$4930 \AA, exhibiting a comparatively lower equivalent velocity of $\sim$13000 km s$^{-1}$.

\begin{table*}
\small{
 \begin{center}
  \begin{threeparttable}
    \caption{Log of the spectroscopic observations of SN~2020ank.}
    \label{tab:tablespec}
    \addtolength{\tabcolsep}{6.5pt}
    \begin{tabular}{ccccccccc} 

    \hline \hline

      Date & MJD &  Phase$^a$ & Instrument & Wavelength & Resolution & Exposure time  & Telescope \\

      (UT) &  & (d) &  & (\AA) & (\AA) & (s) &   \\

      \hline

       2020 Feb 13 & 58,892.926 & $-$\,1.084 & OSIRIS & 3660--7890   & 2.1 & 900  & GTC-10.4m$^b$ \\
       
       2020 Feb 14 & 58,893.360 & $-$\,0.737 &  DBSP  & 3400--10,500 & 1.5 & 500 & P200-5.1m$^b$ \\
       
       2020 Feb 19 & 58,898.000 & +2.980     &  HFOSC & 3500--7800  & 8 & 2700     & HCT-2.0m \\
       
       2020 Feb 20 & 58,899.000 & +3.781     &  HFOSC & 3500--7800  & 8 & 3600     & HCT-2.0m \\
       
       2020 Mar 02 & 58,910.000 & +12.591    &  HFOSC & 3500--7800  & 8 & 2700     & HCT-2.0m \\
    
       2020 Mar 03 & 58,911.000 & +13.392    &  HFOSC & 3500--7800  & 8 & 3600     & HCT-2.0m \\
       \hline 
    \end{tabular}
 \begin{tablenotes}[para,flushleft]
        $^a$ Phase is given in the rest-frame d since \textit{g}-band maximum.
        $^b$ Downloaded from the TNS; \url{https://wis-tns.weizmann.ac.il/object/2020ank}.
    \end{tablenotes}

  \end{threeparttable}
 \end{center}}
\end{table*}

\begin{figure}
\includegraphics[angle=0,scale=0.67]{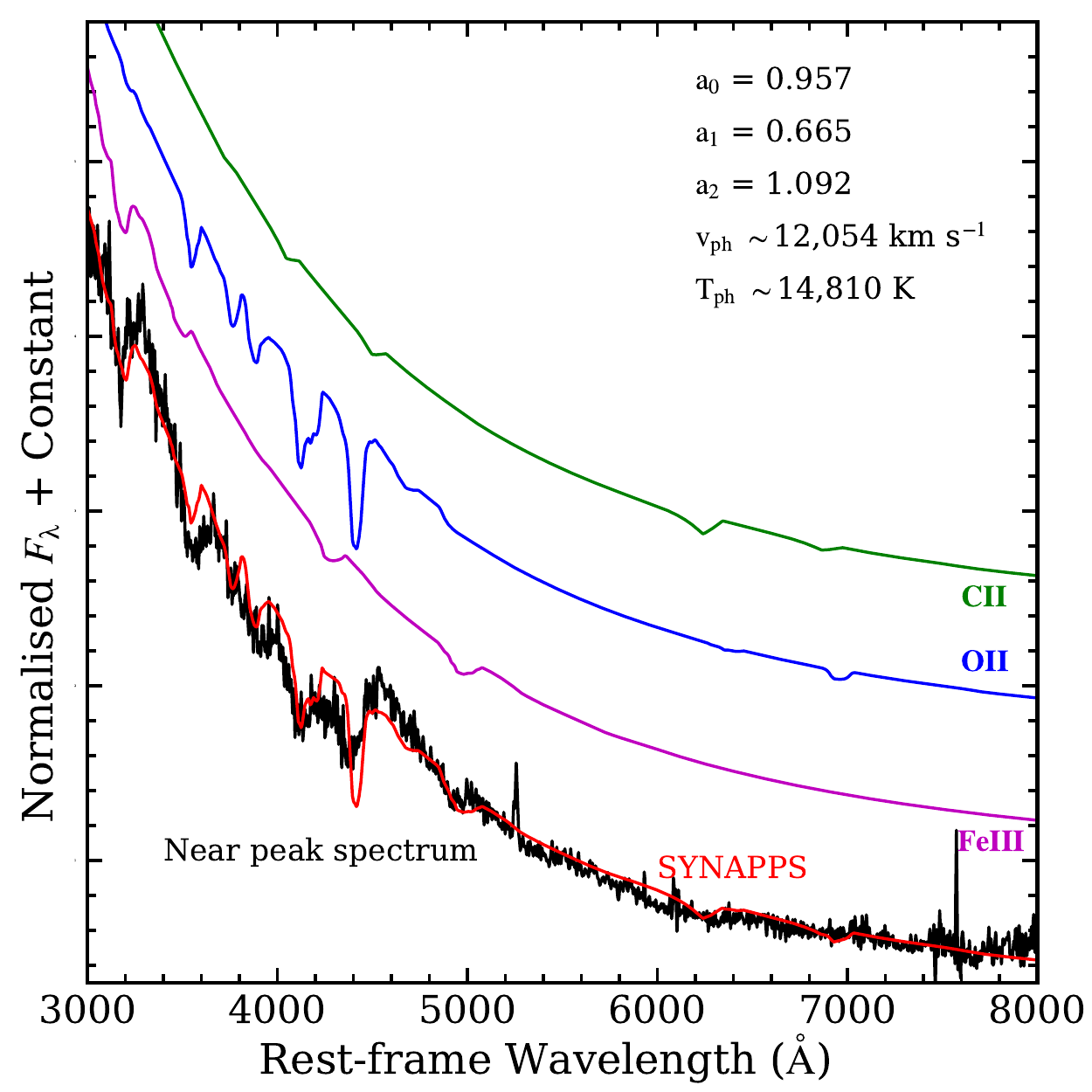}
\caption{The near-peak spectrum of SN~2020ank and the synthetic spectrum (in red) obtained by the {\tt SYNAPPS} \citep{Thomas2011} spectral fitting are shown in the bottom. Individual ion contributions for all the elements that were used to regenerate the best-fitting spectrum are also shown. All the prominent spectral features are well generated by the code using C\,{\sc ii}, O\,{\sc ii}, and Fe\,{\sc iii} elements. The bluer region (from $\sim$3400 to 4500 \AA) of the spectrum is dominated by the W-shaped O\,{\sc ii} features. The C\,{\sc ii} and O\,{\sc ii} features are at higher velocities ($\sim$16500 and 17750 km s$^{-1}$, respectively) in comparison to Fe\,{\sc iii} line velocity ($\sim$13100 km s$^{-1}$).}
\label{fig:indi}
\end{figure}

\subsection{The {\tt SYNAPPS} spectral modelling}\label{sec:synapps}
The parametrized spectrum synthesis code {\tt SYNAPPS} \citep{Thomas2011} is a rewritten, automated, and improved version of the SYNOW code \citep[written in C++;][]{Fisher2000,Branch2002}. The basic assumptions of the code \citep{Thomas2011} are the following: spherical symmetry, homologous expansion of the ejecta, Sobolev approximation for line formation \citep{Sobolev1957}, and considering the ejecta as a gas with the rapid flow. This code's limitations are sharp photosphere, BB assumption, and the concept of no electron scattering \citep{Thomas2011}. The code ignores the effects of continuous opacity and deals with local thermodynamic equilibrium for maintaining populations at different levels. The {\tt SYNAPPS} code uses well-structured input known as a YAML file consisting of different global and local parameters. The global parameters are the photospheric velocity ($v_{ph}$), the outer velocity of the line-forming regions ($v_{out}$), BB photosphere temperature ($T_{ph}$), and the coefficients of quadratic warping function ($a_0$, $a_1$, and $a_2$), which are applied to the synthetic spectrum to match the observed ones. Global parameters decide the overall shape of the spectrum. On the other hand, local parameters like the line opacity ($\tau$), lower and upper cutoff velocities ($v_{min}$ and $v_{max}$, respectively), aux parameter ($aux$), and the Boltzmann excitation temperature ($T$) depend on the individual profiles of the different elements. 

In SN ejecta, matter density in shells above the photosphere could be described by two different laws. The first one is the power law with index ``n'' ($\tau  \propto  v^{-n}$), which is set for all compositions of ions in the model. The second one is the exponential law with the parameter e-folding velocity ``$v_e$'' ($\tau \propto exp(-v/v_e )$), which could be changed for each ion independently. We checked both the cases for our analysis and found that the exponential law is more reliable for fine-tuning. The {\tt SYNAPPS} code can calculate two shapes of the line profile. The first case corresponds to a layer un-detached from the photosphere, giving rise to conspicuous emission and more gentle sloping to the absorption part of the profile. The second case (detached) describes a layer that has a larger $v_{\rm exp}$ than the photosphere. Therefore, there is a gap between the photosphere and successive layers. In our case, the lines arising due to the O\,{\sc ii}, C\,{\sc ii}, and Fe\,{\sc iii} elements are considered as detached from the photosphere as they exhibit higher velocities in comparison to $v_{ph}$. In summary, we suppose that the real distribution of matter densities in these layers could enormously differ from the model conceptions but could be used to describe the more conspicuous and dense part of layers.

\begin{table*}
\small{
  \begin{center}
    \caption{Best-fitting local parameters obtained using the {\tt SYNAPPS} spectral fitting on the near-peak spectrum of SN~2020ank along with the inferred values of $g$, $z(T)$, $n_l$, $N$ and $\rho$ are listed.}
    \label{tab:para_SYNAPPS}
    \addtolength{\tabcolsep}{5pt}
    \begin{tabular}{c c c c c c c c c c c} 

    \hline \hline

    Element  & log\,$\tau$ & $v_{\rm min}$ & $v_{\rm max}$ & $aux$ & $T$ & $g$ & $z(T)$ & log($n_l$) & log(N) &log($\rho$)\\ 
       $ $     &  $ $ & ($10^3$ km s$^{-1}$) & ($10^3$ km s$^{-1}$) &  & ($10^3$ K) & & & cm$^{-3}$ & cm$^{-3}$  & g cm$^{-3}$ \\ 
    \hline
C\,{\sc ii}  &  $-0.773$ & 16.500 & 47.511 & 4.068 & 16.000 & 6 & 6.24 & 3.413  & 5.902 & $-16.798$ \\

O\,{\sc ii}  &  4.554 & 17.740 & 31.934 & 0.651 & 12.522 & --- & --- & --- & ---  & --- \\

Fe\,{\sc iii} &   2.841 & 13.110 & 36.871 & 0.435 & 13.234  & 7 & 34.48 & 10.002 & 12.057  & $-9.975$ \\

      \hline 
    \end{tabular}
  \end{center}}
\end{table*}

A better signal-to-noise ratio of the spectra taken using the GTC-10.4m and P200-5.1m provided an opportunity to perform the {\tt SYNAPPS} spectral modelling. Consequently, we used the {\tt SYNAPPS} \citep{Thomas2011} code to attempt the spectral modelling of the near-peak spectrum (at $-$0.74 d since $g$-band maximum, whereas $\sim$28.4 d since explosion) of SN~2020ank. We were unable to perform it for the post-peak spectra taken with the HCT-2.0m, because of the poor signal-to-noise ratio. We also present the contributions from the individual ions to reproduce the near-peak spectrum of SN~2020ank using the {\tt SYN++} \citep{Thomas2011} and the output obtained from the {\tt SYNAPPS} code, see Fig.~\ref{fig:indi}. All the prominent features are well reproduced by the code. The local fitting parameters (log\,$\tau$, $v_{\rm min}$, $v_{\rm max}$, $aux$, and $T$) for individual ions obtained from the {\tt SYNAPPS} spectral fitting are tabulated in Table~\ref{tab:para_SYNAPPS}. The global parameters ($a_0$, $a_1$, $a_2$, $v_{ph}$, and $T_{ph})$ are presented in upper-right side of Fig.~\ref{fig:indi} itself. The synthetic spectrum is obtained using the C\,{\sc ii}, O\,{\sc ii}, and Fe\,{\sc iii} ions, however, the bluer region (from $\sim3400$ to 4500 \AA) is mainly dominated by the W-shaped O\,{\sc ii} features. The {\tt SYNAPPS} spectral fitting also shows a clear blending of C\,{\sc ii} ($\lambda\lambda$3921 and 4267) and Fe\,{\sc iii} ($\lambda$4430) with the O\,{\sc ii} features in the bluer part of the spectrum. From the best-fitted spectrum, a $T_{ph}$ of $\sim$14810 K and the $v_{ph}$ of $\sim$12050 km s$^{-1}$ are obtained. It is noteworthy that the estimated $v_{ph}$ is consistent within the error bars with the $v_{exp}$ $\sim$12270 $\pm$ 900 km s$^{-1}$ obtained from the bolometric light-curve modelling using the {\tt MINIM} code (see Section~\ref{sec:MINIM}). Also, $T_{ph}$ obtained through the {\tt SYNAPPS} spectral fitting ($\sim$14810 K) closely matched to those independently derived using BB fitting to the spectrum ($\sim$14800 $\pm$ 300 K) and BB fitting to the photometric SED ($\sim$15400 $\pm$ 800 K) at similar epoch. The absorption minima of lines interpreted as C\,{\sc ii} and O\,{\sc ii} were fitted by larger velocities ($\sim$16500 and 17750 km s$^{-1}$, respectively) than for Fe\,{\sc iii} ($\sim$13100 km s$^{-1}$), see Table~\ref{tab:para_SYNAPPS}. However, this spectrum of SN~2020ank could also be modelled using O\,{\sc iii} and C\,{\sc iii} in place of O\,{\sc ii} and C\,{\sc ii} \citep{Hatano1999} as suggested by \citet{Konyves-Toth2020} for the fast-evolving SLSN~2019neq. Using the equations 4 and 6 of \citet{Konyves-Toth2020}, we also estimated the number density of ions at the lower level of the transition ($n_l$), full number density of the ion ($N$), and the mass density ($\rho$, in eV) values for the C\,{\sc ii} and Fe\,{\sc iii} ions (see Table~\ref{tab:para_SYNAPPS}). To calculate the same, we obtained the statistical weight of the lower level ($g$) and the partition function ($z(T)$) from the National Institute of Standards and Technology\footnote{\url{https://www.nist.gov/pml/atomic-spectra-database}} and also tabulated these values in Table~\ref{tab:para_SYNAPPS}. The oscillator strength $(f)$ and the excitation potential of the lower level ($\chi$) are adopted from \citet{Hatano1999}, whereas values of $\tau$ and $T$ are taken from the {\tt SYNAPPS} spectral fitting. We are unable to calculate the values of $n_l$, $N$, and $\rho$ for O\,{\sc ii} ion because the reference line for O\,{\sc ii} ion is considered as a forbidden transition by \cite{Hatano1999}, as also stated by \citet{Konyves-Toth2020}. 

Using the estimated $v_{ph}$ from the {\tt SYNAPPS} spectral fitting and a rise time ($t_r$, time since date of explosion to the peak bolometric luminosity) of $\sim$27.9 $\pm$ 1.0 d (rest-frame), we computed photospheric radius ($r_{phot}$) of $\sim$2.4 $\times$ 10$^{15}$ cm and total optical depth ($\tau_{tot}$) of $\sim$74.6 for SN~2020ank. The details about methods of calculation of the above discussed parameters are well described in \citet{Konyves-Toth2020}. In the case of SN~2020ank, the value of $r_{phot}$ is lower and $\tau_{tot}$ is higher in comparison to those estimated for slow-evolving SN~2010kd ($r_{phot}$ $\sim$6.2 $\times$ 10$^{15}$ cm and $\tau_{tot}$ $\sim$60) and fast-evolving SN~2019neq ($r_{phot}$ $\sim$5.1 $\times$ 10$^{15}$ cm and $\tau_{tot}$ $\sim$43) by \citet{Konyves-Toth2020}. Using the above discussed values of $v_{ph}$ and $t_r$, we also estimated the $M_{ej}$ ($\sim$7.2 $\pm$ 0.5 $M_{\odot}$) and kinetic energy [$E_k$ $\sim$(6.3 $\pm$ 0.1) $\times$ 10$^{51}$ erg] for SN~2020ank applying equations 1 and 3 of \citet{Wheeler2015}. The estimated value of the $E_k$ for SN~2020ank is nearly 2 to 3 times higher than what neutrino driven explosion can give at maximum and favours a jet feedback mechanism \citep{Soker2016, Soker2017}. The calculated $M_{ej}$ $\sim$7.2 $\pm$ 0.5 $M_{\odot}$ of SN~2020ank (using the spectral analysis) is higher than the $M_{ej}$ obtained from the semi-analytical light-curve modelling ($\sim$3.58 $\pm$ 0.04 $M_{\odot}$) using the {\tt MINIM} code (see Section~\ref{sec:MINIM}). Comparatively lower value of $M_{ej}$ obtained using the light-curve modelling might attribute to the underestimation of the peak luminosity by the MAG model. The estimated $M_{ej}$ value of SN~2020ank is lower in comparison to the well-studied slow-evolving PTF12dam \citep[$\sim 10-16 M_\odot$;][]{Nicholl2013}, whereas closer to the fast-evolving SN~2010gx \citep[$\sim 7.1 M_\odot$;][]{Inserra2013}.

\begin{figure*}
\includegraphics[angle=0,scale=1.0]{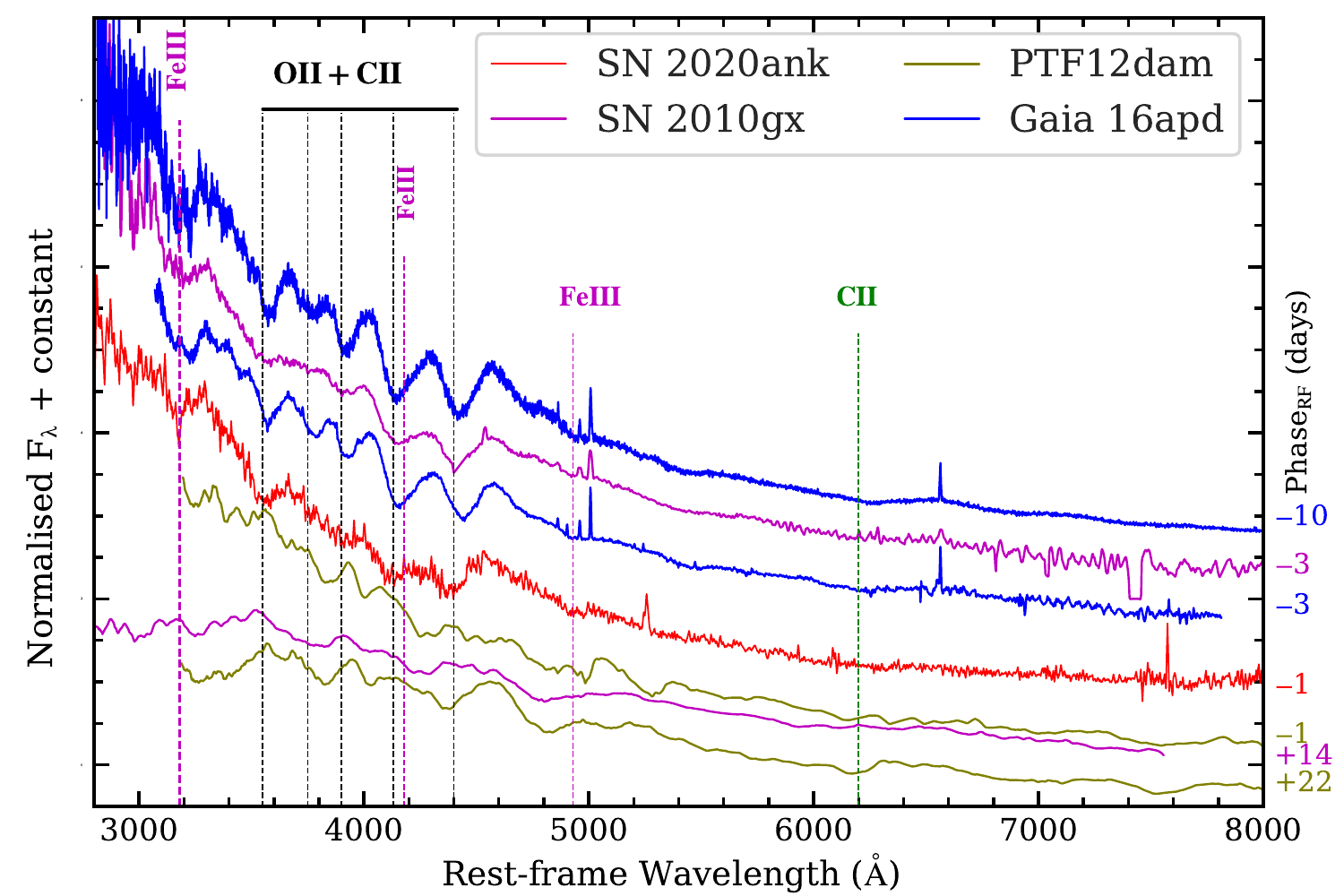}
\caption{The near-peak spectrum of SN~2020ank (in red) is compared with some of the well-studied SLSNe~I: SN~2010gx \citep[in magenta;][]{Pastorello2010}, PTF12dam \citep[in olive;][]{Nicholl2013}, and Gaia16apd \citep[in blue;][]{Kangas2017,Nicholl2017}. The spectral features of SN~2020ank are apparently match with those observed in the case of SN~2010gx suggesting it as yet another fast-evolving SLSN.}
\label{fig:spec_comp}
\end{figure*}

\subsection{Spectral comparison} \label{sec:comp_spectra}

In the section, we compare the near-peak spectrum of SN~2020ank (in red) with the spectra of SN~2010gx \citep[fast-evolving;][]{Pastorello2010}, PTF12dam \citep[slow-evolving;][]{Nicholl2013}, and Gaia16apd \citep[intermediate-decaying;][]{Kangas2017, Nicholl2017}, see Fig.~\ref{fig:spec_comp}. Overall, the spectral features of SN~2020ank are similar to the three other SLSNe~I. The near-peak spectrum of SN~2020ank appears to closely match with the spectrum of SN~2010gx at $-3$ d. The closely matched absorption features of SN~2020ank with those of SN~2010gx indicate nearly equal ejecta velocities. The spectral features of SN~2020ank also match well with the $-3$ d spectrum of Gaia16apd, however, the minima of the absorption features fall at bluer wavelengths, indicating higher ejecta velocity as compared to Gaia16apd. The close spectral resemblance of SN~2020ank with SN~2010gx puts it in the category of ``SN~2011ke-like'' (fast-evolving) SLSN, as suggested by \citet{Quimby2018}. To confirm this fast-spectral evolution, we compare the near-peak spectrum of SN~2020ank with the near- and post-peak spectra (at +22 d) of PTF12dam (slow-evolving). The near-peak spectral features of SN~2020ank appear similar to that were observed in the spectrum of PTF12dam at +22 d, confirming the comparatively faster spectral evolution of SN~2020ank.

\section{Results and Conclusion}\label{sec:results}

We have presented the early optical photometric and spectroscopic observations of the fast-evolving and bright SLSN~2020ank. The photometric data were obtained in the Bessell $U$, $B$, $V$, $R$, and $I$ bands (from $\sim$+3 to +52 d, rest frame) using the ST-1.04m, HCT-2.0m, and the DOT-3.6m along with the publicly available SDSS $g$ and $r$ bands data (from $\sim-$21 to +36 d, rest frame) from the $ZTF$. The spectral analysis were performed on the post-peak spectra observed with the HCT-2.0m (from $\sim$+3 to +13 d, rest frame) at four epochs. These spectral data were also supplemented by two publicly available spectra obtained from the GTC-10.4m and the P200-5.1m (at $-$1.08 and $-$0.74 d, rest frame). Apart from providing well-calibrated photometric data of this newly discovered bright and fast-evolving SLSN, the main findings of the present analysis are the following:

1) The post-peak decay rate of SN~2020ank in the rest-frame $U$-band is consistent with the theoretical decay rate of $^{56}Ni$ $\rightarrow$ $^{56}Co$ (0.11 mag d$^{-1}$) and turns shallower as we go towards redder bands ($\sim$0.05 mag d$^{-1}$ for $R$-band). Also, the light-curve decay of SN~2020ank seems steeper than other well-studied slow-decaying SLSNe~I.

2) The well-sampled rest-frame $g$-band light curve exhibits a peak-absolute magnitude of $\sim$ $-21.84 \pm 0.10$ mag, indicating SN~2020ank as one of the bright SLSNe~I. The pre-peak rising and post-peak decaying rates of the light curve are similar to other well-studied fast-evolving SLSNe~I (e.g., SN~2010gx, PTF11rks, and SN~2011ke) but comparatively steeper in comparison to the slow-evolving SLSNe~I (e.g., SN~2010kd, PTF12dam, and SN~2015bn). However, the rest-frame $g-r$ colour evolution of SN~2020ank is not consistent with the fast-evolving SLSNe~I and is closer to the slow-evolving ones. 

3) The bolometric light curve of SN~2020ank is symmetric around the peak with $L^{rise}_{max}$/e $\approx$ $L^{fall}_{max}$/e $\approx$ 15 d and takes nearly 27 d to reach the $L_{max}$ of $\sim$4 $\times$ 10$^{44}$ erg s$^{-1}$. The estimated value of the  $L_{max}$ for SN~2020ank is higher than SN~2010gx, SN~2010kd, PTF12dam, SN~2015bn, Gaia16apd, etc., whereas lower in comparison to that were observed for SN~2011ke, PTF13ajg, and ASASSN-15lh. Overall, the light-curve comparison of SN~2020ank with other well-studied slow- and fast-evolving SLSNe~I suggests that it is a fast-evolving SLSN having a high peak brightness. 

4) Semi-analytical light-curve modelling using {\tt MINIM} rules out RD and the CSMI as possible powering mechanisms for SN~2020ank. Our findings suggest a spin-down millisecond magnetar having $P_i$ of $\sim$2.23 $\pm$ 0.51 ms and $B$ $\sim$(2.91 $\pm$ 0.07) $\times$ $10^{14}$ G as possibly powering source with total ejected mass of $\sim$3.58 $\pm$ 0.04 $M_{\odot}$. The observed excess $UV$ flux near the peak in the case of SN~2020ank also supports the central engine based power source.

5) Spectroscopic analysis helped to probe chemical composition of the ejecta and constrain crucial physical parameters of SN~2020ank. The {\tt SYNAPPS} spectral modelling reveals that the near-peak spectrum of SN~2020ank is dominated by the W-shaped O\,{\sc ii} features along with comparatively fewer contributions from the C\,{\sc ii} and Fe\,{\sc iii} species. The $T_{ph}$ ($\sim$14800 K) and $v_{ph}$ ($\sim$12050 km s$^{-1}$) estimated by the {\tt SYNAPPS} code are consistent within error bars to the $T_{BB}$ calculated from SED fitting to the photometric data ($\sim$15400 $\pm$ 800 K) and $v_{exp}$ ($\sim$12270 $\pm$ 900 km s$^{-1}$) computed by the best fitted MAG model. 

6) Using the value of $v_{ph}$ obtained from the {\tt SYNAPPS} spectral fitting and assuming diffusion time-scale $\approx$ $t_r$ ($\sim$27.9 $\pm$ 1.0 d) with $\kappa$ = 0.1 cm$^2$ g$^{-1}$, we constrain the $r_{phot}$ ($\sim$2.4 $\times$ 10$^{15}$ cm), $\tau_{tot}$ ($\sim$74.6), $M_{ej}$ ($\sim$7.2 $M_{\odot}$), and $E_k$ ($\sim$6.3 $\times$ 10$^{51}$ erg) for SN~2020ank.

7) The near-peak spectral comparison of SN~2020ank shows that the spectral features are similar to those observed in SN~2010gx and Gaia16apd. However, higher blueshifted absorption features in SN~2020ank than Gaia16apd indicate higher ejecta velocity in the former. The near-peak spectral similarity with SN~2010gx and apparent similarity with the spectrum of the PTF12dam (slow-evolving) at +22 d confirms comparatively faster spectral evolution of SN~2020ank.

\section*{Acknowledgements}
\addcontentsline{toc}{section}{Acknowledgements}
This study uses the data from the ST-1.04m, HCT-2.0m, and DOT-3.6m, and authors of this paper are highly thankful to the observing staff and observing assistants for their support during observations of SN~2020ank in the trying times of the COVID-19 pandemic. AK and SBP acknowledge the DOT-3.6m proposal no. DOT-2020-C2-P42 for obtaining the optical data of this event. AK and SBP also acknowledge the guidance and discussions related to the physics of core-collapse Supernovae with Prof. J. Craig Wheeler and Prof. Jozsef Vink{\'o}. AK would like to thank Prof. Matt Nicholl, Raya Dastidar, and Kaushal Sharma for the insightful discussion. AK and team members thank the anonymous referee for his/her constructive comments that helped to improve the manuscript. This study also uses the publicly available photometric data of the Samuel Oschin 48-inch Telescope at the Palomar Observatory as part of the $ZTF$ project. $ZTF$ is supported by the National Science Foundation under grant no. AST-1440341 and a collaboration including Caltech, IPAC, the Weizmann Institute for Science, the Oskar Klein Center at Stockholm University, the University of Maryland, the University of Washington, Deutsches Elektronen-Synchrotron and Humboldt University, Los Alamos National Laboratories, the TANGO Consortium of Taiwan, the University of Wisconsin at Milwaukee and Lawrence Berkeley National Laboratories. Operations are conducted by COO, IPAC, and UW. We used the publicly available spectrum of SN~2020ank based on observations made with the GTC-10.4m, installed at the Spanish Observatorio del Roque de los Muchachos of the Instituto de Astrofísica de Canarias, in the island of La Palma. This study also used the publicly available spectrum of P200 Hale telescope. SBP, KM, AA, and RG acknowledge BRICS grant DST/IMRCD/BRICS/Pilotcall/ProFCheap/2017(G) and DST/JSPS grant DST/INT/JSPS/P/281/2018 for this work. This research has utilized the NED, which is operated by the Jet Propulsion Laboratory, California Institute of Technology, under contract with NASA. We acknowledge the use of NASA's Astrophysics Data System Bibliographic Services. This research also made use of the Open Supernova Catalog (OSC) currently maintained by James Guillochon and Jerod Parrent.

\section*{Data Availability}

The photometric and spectroscopic data used in this work can be made available on request to the corresponding authors.

\bibliographystyle{mnras}
\bibliography{2020ank}
\label{lastpage}
\end{document}